\documentclass[sigconf]{acmart}
\usepackage{amsmath,amssymb,amsfonts}
\usepackage{pifont}
\usepackage{graphicx}
\usepackage{textcomp}
\usepackage{lipsum}
\usepackage{booktabs}
\usepackage{tabularx}%%},
\usepackage{framed}
\usepackage{adjustbox}
\usepackage{comment}
\usepackage{multirow}
\usepackage{booktabs, makecell}
\usepackage[ruled]{algorithm2e}
\usepackage{enumerate}
\usepackage{algorithmic}
\usepackage{threeparttable}
\usepackage{subfigure}
\usepackage{caption}%,subcaption}
\usepackage{framed} 
\usepackage{listings}
\usepackage{tcolorbox}
\usepackage{xcolor,color,colortbl}

\usepackage{caption}

\newcommand{\distance}{0pt}
\newcommand{\hhline}{\Xhline{1pt}}
\newcommand{\parabf}[1]{\noindent\textbf{#1}}
\newcommand{\codeIn}[1]{\texttt{#1}}
\newcommand{\numOldDefect}{395}

\newcommand{\mybox}{-2pt}

\newcommand{\tOnesbfl}{117}

\newcommand{\tOneRep}{161}
\newcommand{\tOneMut}{141}

\newcommand{\fulltime}{Time$_{f}$}
\newcommand{\parttime}{Time$_{p}$}
\newcommand{\ourtool}{ProFL}

\newcommand{\prog}{\mathcal{P}_o}
\newcommand{\tsts}{\mathcal{T}}
\newcommand{\tst}{t}
\newcommand{\stmts}{\mathcal{S}}
\newcommand{\stmt}{s}
\newcommand{\meths}{\mathcal{E}}
\newcommand{\meth}{e}
\newcommand{\bugs}{\mathbb{B}}
\newcommand{\patches}{\mathbb{P}}
\newcommand{\patch}{\mathcal{P}}
\newcommand{\groupi}{\mathcal{G}_i}
\newcommand{\pMatrix}{\mathbb{M}}
\newcommand{\pMatrixpart}{\mathbb{M}_p}
\newcommand{\pMatrixfull}{\mathbb{M}_f}
\newcommand{\pMatrixpartCell}[2]{\mathbb{M}_p[#1,#2]}
\newcommand{\pMatrixCell}[2]{\mathbb{M}[#1,#2]}
\newcommand{\unknown}{\ding{109}}%
\newcommand{\pass}{\ding{51}}%
\newcommand{\fail}{\ding{55}}%

\newcommand{\numsbfl}{34}

\newcommand{\susp}[1]{\mathbb{S}[#1]}
\newcommand{\group}[1]{\mathbb{G}[#1]}

\newcommand{\patchLoc}[1]{\mathbb{P}[#1]}
\newcommand{\prelation}[1]{\mathbb{R}[#1]}
\newcommand{\patchgroup}{patch group} % 4 big groups
\newcommand{\susplevel}{suspicious level} % 
\newcommand{\patchoneL}{Clean-Fix Patch}  % long name 
\newcommand{\patchtwoL}{Noisy-Fix Patch} 
\newcommand{\patchthreeL}{None-Fix Patch} 
\newcommand{\patchfourL}{Negative-Fix Patch} 
\newcommand{\patchoneS}{\texttt{CleanFix}} % short name 
\newcommand{\patchtwoS}{\texttt{NoisyFix}} 
\newcommand{\patchthreeS}{\texttt{NoneFix}} 
\newcommand{\patchfourS}{\texttt{NegFix}} 
\newcommand{\patchone}{$Patch_{Clean}$} % short name 
\newcommand{\patchtwo}{$Patch_{Noisy}$} 
\newcommand{\patchthree}{$Patch_{None}$} 
\newcommand{\patchfour}{$Patch_{Neg}$}

\newcommand{\rank}{\succeq}
\newcommand{\smallpatchoneL}{Clean-All-Fix Patch } % short name 
\newcommand{\smallpatchtwoL}{Clean-Part-Fix Patch}
\newcommand{\smallpatchthreeL}{Noise-All-Fix Patch}
\newcommand{\smallpatchfourL}{Noise-Part-Fix Patch}
\newcommand{\smallpatchoneS}{\texttt{CleanAllFix}} % short name 
\newcommand{\smallpatchtwoS}{\texttt{CleanPartFix}}
\newcommand{\smallpatchthreeS}{\texttt{NoisyAllFix}}
\newcommand{\smallpatchfourS}{\texttt{NoisyPartFix}}

\newcommand{\apbasic}{$Basic$} 
\newcommand{\varone}{\ourtool$_{R_1}$} 
\newcommand{\vartwo}{\ourtool$_{R_2}$} 
\newcommand{\varthree}{\ourtool$_{R_3}$} 
\newcommand{\varfour}{\ourtool$_{R_4}$}

\newcommand{\fp}{$f2p$} 
\newcommand{\pf}{$p2f$}

\newcommand{\executionorderone}{$O_1$} 
\newcommand{\executionordertwo}{$O_2$} 
\newcommand{\executionorderthree}{$O_3$} 

\newcommand{\orderone}{$\mathbb{M}_p^{(O_1)}$} 
\newcommand{\ordertwo}{$\mathbb{M}_p^{(O_2)}$} 
\newcommand{\orderthree}{$\mathbb{M}_p^{(O_3)}$}

\newcommand{\bugone}{$Bug_{full}$} 
\newcommand{\bugtwo}{$Bug{part}$} 
\newcommand{\bugthree}{$Bug_{none}$} 

\newcommand{\topone}{Top-1}
\newcommand{\topthree}{Top-3} 
\newcommand{\topfive}{Top-5} 
\newcommand{\topn}{Top-N}
\newcommand{\mar}{MAR}
\newcommand{\mfr}{MFR}
\newcommand{\faultratio}{Ratio_b}

\newcommand{\bugreveal}{bug revealing}

\newcommand{\prapr}{PraPR} 

\newcommand{\muse}{MUSE}
\newcommand{\meta}{Metallaxis}
\newcommand{\pkbasic}{PRFL}
\newcommand{\pkag}{PRFL$_{MA}$}
\newcommand{\dpfl}{DeepFL}
\newcommand{\metallaxis}{Metallaxis}

\newcommand{\fl}{fault localization}
\newcommand{\Fl}{Fault localization}
\newcommand{\mbfl}{MBFL}
\newcommand{\sbfl}{SBFL}
\newcommand{\spe}{\emph{Spectrum}}
\newcommand{\hyb}{MCBFL}

\newcommand{\apr}{APR\xspace}

\newcommand{\fin}{feedback information}

\newcommand{\PraPR}{PraPR}

\newcommand{\olddefects}{{Defects4J (V1.2.0)}}  
\newcommand{\newdefects}{{Defects4J (V1.4.0)}} 

\newcommand{\Comment}[1]{}

\newcommand{\lingming}[1]{\textcolor{red}{[Lingming: #1]}}
\newcommand{\dan}[1]{\textcolor{red}{[dan: #1]}}

\definecolor{GrayOne}{gray}{0.9}
\definecolor{GrayTwo}{gray}{0.8}
\definecolor{GrayThree}{gray}{0.7}
\definecolor{GrayFour}{gray}{0.6}
\definecolor{GrayFive}{gray}{0.5}

\newcommand{\defects}{Defects4J}

\newcommand{\naive}{na\"ive\xspace}

\newcommand{\exeOverheadGain}{96.2\% \xspace}
\newcommand{\mfrImrovement}{33.61\%\xspace}

\setcopyright{none}
\renewcommand\footnotetextcopyrightpermission[1]{}

\definecolor{notecolor}{rgb}{0.0, 0.5, 0.69}
\newcommand{\note}[1]{{\scriptsize \textcolor{gray}{#1}}}
\begin{document}

\lstdefinestyle{java}{ % Define a style for your code snippet, multiple definitions can be made if, for example, you wish to insert multiple code snippets using different programming languages into one document
	%    backgroundcolor=\color{highlight}, % Set the background color for the snippet - useful for highlighting
	language=java,
	basicstyle=\scriptsize\ttfamily, % The default font size and style of the code
	breakatwhitespace=false, % If true, only allows line breaks at white space
	breaklines=true, % Automatic line breaking (prevents code from protruding outside the box)
	captionpos=b, % Sets the caption position: b for bottom; t for top
	commentstyle=\color[rgb]{0.0, 0.5, 0.69},%\color[rgb]{0,0.6,0}, % Style of comments within the code - dark green courier font
	deletekeywords={}, % If you want to delete any keywords from the current language separate them by commas
	%escapeinside={\%}, % This allows you to escape to LaTeX using the character in the bracket
	escapeinside={<@}{@>},
	firstnumber=1, % Line numbers begin at line 1
	frame=lines, % Frame around the code box, value can be: none, leftline, topline, bottomline, lines, single, shadowbox
	frameround=tttt, % Rounds the corners of the frame for the top left, top right, bottom left and bottom right positions
	keywordstyle={[1]\color{blue!90!black}},
	keywordstyle={[3]\color{red!80!orange}},
	morekeywords={Calendar}, % Add any functions no included by default here separated by commas
	numbers=none, % Location of line numbers, can take the values of: none, left, right
	numbersep=-8pt, % Distance of line numbers from the code box
	numberstyle=\tiny\color[rgb]{0.1,0.1,0.1}, % Style used for line numbers
	rulecolor=\color{black}, % Frame border color
	showstringspaces=false, % Don't put marks in string spaces
	showtabs=false, % Display tabs in the code as lines
	stepnumber=1, % The step distance between line numbers, i.e. how often will lines be numbered
	tabsize=2, % Number of spaces per tab in the code
	backgroundcolor=\color{white}
}

%\title{Automated Feedback-based Fault Localization}
\title{Can Automated Program Repair Refine Fault Localization?}

\author{Yiling Lou}
\authornote{This work was mainly done when Yiling Lou was a visiting student in UT Dallas.}
\affiliation{%
 \institution{HCST (Peking University), China}
}
\email{louyiling@pku.edu.cn}

\author{Ali Ghanbari}
\author{Xia Li}
\author{Lingming Zhang}
\affiliation{%
 \institution{UT Dallas, USA}
}
\email{{Ali.Ghanbari, Xia.Li3, lingming.zhang}@utdallas.edu}

\author{Dan Hao}
\author{Lu Zhang}
\affiliation{%
\institution{HCST (Peking University), China}
}
\email{{haodan,zhanglucs}@pku.edu.cn}

% The abstract is a short summary of the work to be presented in the article.
\begin{abstract}
Software bugs are prevalent in modern software systems and notoriously hard to debug manually. Therefore, a large body of research efforts have been dedicated to automated software debugging, including both automated fault localization and program repair. However, the existing fault localization techniques are usually ineffective on real-world software systems while even the most advanced program repair techniques can only fix a small ratio of real-world bugs. Although fault localization and program repair are inherently connected, we observe that in the literature their only connection is that program repair techniques usually use off-the-shelf fault localization techniques (e.g., Ochiai) to determine the potential candidate statements/elements for patching. In this work, we explore their connection in the other direction, i.e., \emph{can program repair in turn help with fault localization?} In this way, we not only open a new dimension for more powerful fault localization, but also extend the application scope of program repair to all possible bugs (not only the bugs that can be directly automatically fixed). We have designed \ourtool{}, a simplistic approach using patch-execution results (from program repair) as the feedback information for fault localization. The experimental results on the widely used \defects{} benchmark show that the basic \ourtool{} can already localize \tOneRep{} of the \numOldDefect{} studied bugs within \topone{}, while state-of-the-art spectrum and mutation based fault localization techniques at most localize \tOnesbfl{} within \topone{}. We also demonstrate \ourtool{}'s effectiveness under different settings. Lastly, we show that \ourtool{} can further boost state-of-the-art fault localization via both unsupervised and supervised learning. 

\end{abstract}
\maketitle

\section{Introduction}
%software bugs are prevalent and challenging to debug
Software bugs (also called software faults, errors, defects, flaws, or failures~\cite{bug}) are prevalent in modern software systems, and have been widely recognized as notoriously costly and disastrous.\Comment{In 2002,
the US National Institute of Standards and Technology (NIST) estimated
that the US economy lost \$60 Billion each year due to software
bugs~\cite{tassey2002economic}. In 2013, a study by Cambridge
University stated that the global cost of software testing/debugging had risen to \$312 Billion
annually~\cite{britton2013reversible}.} For example, in 2017, {\tt
Tricentis.com} investigated software failures impacting 3.7 Billion
users and \$1.7 Trillion assets, and reported that this is
just scratching the surface -- \emph{there can be far more software
bugs in the world than we will likely ever know
about}~\cite{tricentis}.
In practice, software debugging is widely adopted for removing software bugs\Comment{ and ensuring software quality}. However, manual debugging can be extremely tedious, challenging,
and time-consuming due to the increasing complexity of modern software
systems~\cite{tassey2002economic}\Comment{ (e.g., Google has over 2 Billion lines of code in a single
repository~\cite{potvin2016google})}.\Comment{ It has been reported that debugging can take up to 80\% of the total software
cost~\cite{tassey2002economic}.} Therefore, a large body of research efforts have
been dedicated to automated debugging to reduce manual-debugging
efforts~\cite{myers2011art,
perry2007effective, kit1995software,
bertolino2007software,tassey2002economic}.

%automated software debugging and their problems
There are two key questions in software debugging: (1) {\em how to automatically
localize software bugs to facilitate manual repair?} (2) {\em how to automatically
repair software bugs without human intervention?} To address them, researchers have proposed two categories of techniques, \emph{fault localization
}~\cite{zhang2013injecting,moon2014ask,abreu2007accuracy,dallmeier2005lightweight,jones2002visualization,xuan2014test,liblit2005scalable\Comment{,briand2007using,roychowdhury2011novel,roychowdhury2011software,roychowdhury2012family}} and \emph{program
repair}~\cite{bib:GNFW12,
bib:LR15,weimer2013leveraging, bib:LR16b,\Comment{
gao2015fixing,gao2015safe,mechtaev2015directfix,bib:MYR16,demarco2014automatic,
jifeng2016,wei2010automated,
pei2011code,demsky2006inference,jin2011automated,
bib:Weim06,}bib:QMLDZW14,
kong2015experience,qi2015analysis,kim2013automatic}. For example, pioneering spectrum-based fault localization (\sbfl) techniques~\cite{jones2002visualization, abreu2007accuracy, dallmeier2005lightweight} compute the code elements covered by more failed tests or less passed tests as more suspicious, pioneering mutation-based fault localization (\mbfl) techniques~\cite{papadakis2015metallaxis, zhang2013injecting, moon2014ask} inject code changes (e.g., changing \codeIn{>} into \codeIn{>=}) based on mutation testing~\cite{DBLP:journals/tse/JiaH11,DBLP:journals/computer/DeMilloLS78} to each code element to check its impact on test outcomes, and pioneering search-based program repair techniques (e.g., GenProg~\cite{bib:GNFW12}) tentatively change program elements based on certain rules (e.g., deleting/changing/adding program elements) and use the original test suite as the oracle to validate the generated patches. Please refer to the recent surveys on automated software debugging for more details~\cite{bib:WGLW16,bib:Monp18}. 
To date, unfortunately, although debugging has been extensively studied and even has drawn attention from industry (e.g., FaceBook~\cite{marginean2019sapfix,scott2019getafix} and Fujitsu~\cite{saha2017elixir}),
\emph{we still lack practical automated debugging techniques}: (1) existing fault localization techniques have been shown to have limited
effectiveness in practice~\cite{parnin2011automated,xie2016revisit}; (2) existing program repair techniques can only fix a small ratio of real bugs~\cite{jiang2018shaping,DBLP:conf/issta/GhanbariBZ19, bib:WCWHC18} or specific types of bugs~\cite{marginean2019sapfix}.

%our insight
In this work, we aim to revisit the connection between program repair and fault localization for more powerful debugging.
We observe that the current existing connection between fault localization and program repair is that program repair techniques usually use off-the-shelf fault localization techniques to identify potential buggy locations for patching, e.g., the Ochiai~\cite{abreu2007accuracy} \sbfl{} technique is leveraged in many recent program repair techniques, including \prapr~\cite{DBLP:conf/issta/GhanbariBZ19}, CapGen~\cite{bib:WCWHC18}, and SimFix~\cite{jiang2018shaping}. Different from prior work, we aim to connect program repair and fault localization in the reversed way, and explore the following question, \emph{can program repair in turn help with fault localization?} Our basic insight is that the patch execution information during program repair can provide useful feedbacks and guidelines for powerful fault localization. Based on this insight, we designed, \ourtool{} (\textbf{Pro}gram Repair for \textbf{F}ault \textbf{L}ocalization), a simplistic feedback-driven fault localization approach that leverages patch-execution information from state-of-the-art \prapr~\cite{DBLP:conf/issta/GhanbariBZ19} repair tool for rearranging fault localization results computed by off-the-shelf fault localization techniques. Note that even state-of-the-art program repair techniques can only fix a small ratio of real bugs (i.e., <20\% for \defects~\cite{DBLP:conf/issta/GhanbariBZ19, bib:WCWHC18, jiang2018shaping}) fully automatically and were simply aborted for the vast majority of unfixed bugs, while our approach extends the application scope of program repair to all possible bugs -- \emph{program repair techniques can also provide useful fault localization information to help with manual repair even for the bugs that are hard to fix automatically.}

%our study
We have evaluated our \ourtool{} on the \olddefects{} benchmark, which includes \numOldDefect{} real-world bugs from six open-source Java projects and has been widely used for evaluating both fault localization and program repair techniques~\cite{bib:WCWHC18,li2017transforming, sohn2017fluccs, jiang2018shaping, DBLP:conf/issta/GhanbariBZ19}. Our experimental results show that \ourtool{} can localize \tOneRep{} bugs within \topone{}, while state-of-the-art spectrum and mutation based fault localization techniques can at most localize \tOnesbfl{} bugs within \topone. We further investigate the impacts of various experimental configurations: (1) we investigate the finer-grained patch categorizations and observe that they do not have clear impact on \ourtool{}; (2) we investigate the impact of different off-the-shelf \sbfl{} formulae used in \ourtool{}, and observe that \ourtool{} consistently outperforms traditional \sbfl{} regardless of the used formulae; (3) we replace the repair feedback information with traditional mutation feedback information in \ourtool{} (since they both record the impacts of certain changes to test outcomes), and observe that \ourtool{} still localizes \tOneMut{} bugs within \topone{}, significantly outperforming state-of-the-art \sbfl{} and \mbfl; (4) we feed \ourtool{} with only \emph{partial} patch-execution information (since the test execution will be aborted for a patch as soon as it gets falsified by some test for the sake of efficiency in practical program repair scenario), and observe that, surprisingly, \ourtool{} using such partial information can reduce the execution overhead by \exeOverheadGain with negligible effectiveness drop; (5) we also apply \ourtool{} on a newer version of Defects4J, \newdefects~\cite{newdefects4j}, and observe that \ourtool{} performs consistently. In addition, we further observe that \ourtool{} can even significantly boost state-of-the-art fault localization via both unsupervised~\cite{zhang2017boosting, zhang2019empirical} and supervised~\cite{li2019deepfl} learning, localizing 185 and 216.8 bugs within \topone{}, the best fault localization results via unsupervised/supervised learning to our knowledge.

\Comment{First, note that traditional mutation testing~\cite{mutation-papers} can also change the program under test in different ways and evaluate the impact of such changes against the original test suite. Therefore, the proposed techniques can also use traditional mutation information~\cite{mutation-papers} (instead of the program repair information) as the feedback for fault localization (similar with traditional \mbfl). Our experimental results show that even using the same mutation information as \mbfl, our techniques can still significantly outperform the existing \sbfl{} and \mbfl{} techniques. Second, so far, we have forced each patch to be executed by all tests covering it. However, in practical program repair scenario, the test execution will be aborted for a patch as soon as it gets falsified by some test for the sake of efficiency, i.e., only \emph{partial} patch-execution information will be traced.} 

%contribution
This paper makes the following contributions:
\begin{itemize}
    \item This paper opens a new dimension for improving fault localization via off-the-shelf program repair techniques, and also extends the application scope of program repair techniques to all possible bugs (not only the bugs that can be directly automatically fixed).
    \item We have implemented a fully automated feedback-driven fault localization approach, \ourtool, based on the patch-execution results from state-of-the-art program repair technique, \prapr.
    \item We have performed an extensive study of the proposed approach on the widely used \defects{} benchmarks, and demonstrated the effectiveness, efficiency, robustness, and general applicability of the proposed approach. \Comment{: (1) simplistic feedback-driven fault localization based on program repair information can significantly outperform state-of-the-art spectrum and mutation based fault localization, (2) the proposed approach consistently outperforms corresponding \sbfl{} techniques using different \sbfl{} formulae, (3) the proposed approach still outperforms the baseline techniques when using simple mutation information as feedback, (4) the proposed approach has negligible effectiveness drop when using the partial patch execution information, (5) the proposed approach can further boost state-of-the-art learning-based fault localization, and (6) the proposed approach also performs consistently on additional bugs besides \olddefects{}.}
\end{itemize}
\section{Background and Related Work}

\parabf{Fault Localization}~\cite{pearson2017evaluating, zhang2013injecting,moon2014ask,abreu2007accuracy,
  dallmeier2005lightweight,jones2002visualization,xuan2014test,
  liblit2005scalable,briand2007using,roychowdhury2011novel,roychowdhury2011software,roychowdhury2012family}
aims to precisely diagnose potential buggy locations to facilitate manual bug
fixing. The most widely studied \emph{spectrum-based fault localization}
(\sbfl) techniques usually apply statistical analysis (e.g.,
Tarantula~\cite{jones2002visualization}, Ochiai~\cite{abreu2007accuracy}, and
Ample~\cite{dallmeier2005lightweight}) or learning techniques~\cite{briand2007using, roychowdhury2011novel,
  roychowdhury2011software, roychowdhury2012family} to the execution
traces of both passed and failed tests to identify the most suspicious
code elements (e.g., statements/methods). The insight is that code
elements primarily executed by failed tests are more
suspicious than the elements primarily executed by passed
  tests. However, a code element executed by a failed test does not
necessarily indicate that the element has impact on the test execution
and has caused the test failure.\Comment{ Due to this gap between coverage and impact,
traditional \sbfl{} has been shown to have
limited effectiveness in
practice~\cite{parnin2011automated,xie2016revisit}. } To bridge the gap
between coverage and impact information, researchers proposed {\em
  mutation-based fault localization} (\mbfl)~\cite{moon2014ask,
  papadakis2012using, papadakis2015metallaxis, zhang2013injecting},
which\Comment{ transforms program source code based on mutation testing to
check the impact of each code element on the test outcomes.  As shown
in Section~\ref{sec:proposed}, mutation testing was originally
proposed to evaluate test effectiveness by injecting artificial faults
into the program under
test~\cite{jia2011analysis,hamlet1977testing,demillo1978hints}. In
\mbfl, mutation testing is used to} injects changes to each code element (based on mutation testing~\cite{DBLP:journals/tse/JiaH11,DBLP:journals/computer/DeMilloLS78})
to check its impact on the test outcomes.\Comment{ The basic idea of
\metallaxis{}~\cite{papadakis2012using, papadakis2015metallaxis}, the
first \mbfl{} technique, is that if one mutant incurs different
failure outputs/messages for failed tests, the corresponding code
element of this mutant may have high impact on failed tests, and may
be the cause of the test failures. \metallaxis{} directly applies the
existing spectrum-based formulae to the impact information to compute
code element suspiciousness. The basic idea of
\muse{}~\cite{moon2014ask}, another representative \mbfl{} technique, is that
mutating faulty elements may mask the fault and make some failed tests
pass, while mutating correct elements may lead to more faulty elements
besides existing faulty elements, making more tests fail.\Comment{
Based on this insight, \muse{} uses a newly designed formula for
localizing faulty code elements.}} \mbfl{} has been applied to both general bugs (pioneered by
\metallaxis~\cite{papadakis2012using,papadakis2015metallaxis}) and regression bugs (pioneered by FIFL~\cite{zhang2013injecting}). \ourtool{} shares similar insight with \mbfl{} in that program changes can help determine the impact of code elements on test failures. However, \ourtool{} utilizes program repair information that aims to fix software bugs to pass more tests rather than mutation testing that was originally proposed to create new artificial bugs to fail more tests; \ourtool{} also embodies a new feedback-driven fault localization approach. 
Besides \sbfl{} and \mbfl, researchers have proposed to utilize various other information for fault localization (such as program slicing~\cite{xuan2014test}, predicate switching~\cite{zhang2006locating}, code complexity~\cite{sohn2017fluccs}, and program invariant~\cite{b2016learning} information), and have also utilized supervised learning to incorporate such different feature dimensions for fault localization~\cite{xuan2014learning, li2017transforming, li2019deepfl}. However, the effectiveness of supervised-learning-based fault localization techniques may largely depend on the training sets, which may not always be available. Therefore, researchers recently have also proposed to recompute \sbfl{} suspiciousness by considering the contributions of different tests via the unsupervised-learning-based PageRank analysis~\cite{zhang2017boosting, zhang2019empirical}. In this work, we aim to explore a new direction for simplistic fault localization without supervised learning, i.e., leveraging patch-execution information (from program repair) for powerful fault localization.

\parabf{Automated Program Repair}\Comment{a.k.a. automatic program/software patching or
  automatic fixing. Consult Monperrus' survey \cite{bib:Monp18} to get
  a list of synonyms.} (\apr)
techniques~\cite{\Comment{bib:TKKX14,bib:LDR14,bib:PKLABCPSSSWZER09,bib:GNFW12,
  bib:DW10,bib:JCMY16,bib:TR15,bib:WCWHC18,}bib:LR15,bib:LR16b,bib:NQRC13,
  bib:XMDCMDBM17, bib:LPF17,
  bib:MYR16,bib:PFNWMZ14,bib:DZM09,bib:GMK11,
  bib:GMM17,bib:Monp18,bib:MDSXM17,bib:Weim06} aim to directly fix software bugs
with minimal human intervention via synthesizing \emph{genuine} patches (i.e., the patches
semantically equivalent to developer patches)\Comment{ or suggesting patches
that may guide the debuggers to fix the bugs faster}. Therefore,
despite a young research area, \apr{} has been extensively studied in the literature. \Comment{Based on the actions they take for fixing a bug,}
State-of-the-art \apr techniques can be divided into two broad
categories: (1) techniques that dynamically monitor program executions to find deviations from certain specifications, and then
\emph{heal} the program under test via modifying its runtime states in case of
abnormal behaviors~\cite{bib:LDR14,bib:PKLABCPSSSWZER09}; (2) techniques that directly modify program code
representations based on different rules/strategies, and
then use either tests or formal specifications\Comment{ (such as code
contracts)} as the oracle to validate each generated candidate patch to
find \emph{plausible} patches (i.e., the patches passing all tests/checks)~\cite{\Comment{bib:GNFW12,bib:DW10,bib:JCMY16,bib:TR15,bib:WCWHC18,bib:LR15,}bib:LR16b,bib:NQRC13,bib:XMDCMDBM17,bib:LPF17,bib:MYR16\Comment{},bib:PFNWMZ14,bib:DZM09,bib:GMK11}. Among these code-representation-level techniques, those based on tests have gained popularity since\Comment{ run-time healing techniques are not
applicable in many circumstances\Comment{cons of healing techniques},
and main-stream programming languages lack constructs for supporting
contract-based programming\Comment{cons of contract-based
techniques}. Furthermore,} testing is the prevalent methodology for
detecting software bugs in practice\Comment{, since very few software systems are based on
rigorous, formal specifications}. Based on different hypotheses,
state-of-the-art code-representation-level techniques leverage a variety of strategies to
generate/synthesize patches. \emph{Search-based} \apr techniques\Comment{are based on the hypothesis} assume that most bugs could be solved by
searching through all the potential candidate patches based on certain
patching rules (i.e., program-fixing templates)~\cite{bib:DW10,bib:GNFW12, bib:WCWHC18,
jiang2018shaping}. Alternatively, \emph{semantics-based} techniques use
deeper semantical analyses (including symbolic
execution~\cite{king1976symbolic, clarke1976system}) to synthesize program
conditions, or even more complex code snippets, that can pass all the
tests~\cite{bib:MYR16,bib:NQRC13,bib:XMDCMDBM17}. Recently,
search-based \apr has been extensively studied due to its scalability
on real-world systems, e.g., the most recent \prapr{} technique has been reported to produce genuine patches for 43 real bugs from \defects{}\Comment{ (a set
of real-world Java programs widely used for evaluating \apr techniques}~\cite{bib:JJE14}. Despite the success of recent advanced \apr techniques, even the most recent program repair technique can only fix a small ratio (i.e., <20\% for \defects{})
of real bugs~\cite{jiang2018shaping,DBLP:conf/issta/GhanbariBZ19, bib:WCWHC18} or specific types of bugs~\cite{marginean2019sapfix}. 

In this work, we aim to leverage program repair results to help with fault localization. More specifically, we design, \ourtool, a simplistic feedback-driven fault localization approach guided by patch-execution results (from program repair). Note that the recent work \prapr{}~\cite{DBLP:conf/issta/GhanbariBZ19} has briefly mentioned that plausible patches may potentially help localize bugs. However, it does not present a systematic fault localization approach working for all possible bugs, and was only demonstrated on a small number of bugs with plausible patches. In contrast, we present the first systematic fault localization approach driven by program repair results and perform the first extensive study under various settings and on a large number of real-world bugs. Feedback-driven fault localization techniques have also been investigated before~\cite{DBLP:conf/icse/LinSXLD17,li2018enlightened}. However, existing feedback-driven fault localization techniques usually require manual inspection to guide the debugging process. In contrast, we present a fully automated feedback-driven fault localization, i.e., \ourtool{} utilizes program fixing attempts and corresponding patch-execution results as feedback to enable powerful automated fault localization.

\begin{comment}
\begin{lstlisting}[style=java,numbers=none,label=lst:lang-21-pattern,caption=Referenced history patch to fix Lang-21.]
<@\note{Commit : github.com/Cbsoftware/PressureNet/commit/9d00742}@>
<@\note{Message: Fixing time display bugs, \#113}@>
<@\note{Source : src.ca.cumulonimbus.barometernetwork.BarxxActivity}@>
==========
 - if(start.get(Calendar.HOUR)==0&&end.get(Calendar.HOUR)==0){
 + if(start.get(Calendar.HOUR_OF_DAY)==0&&end.get(Calendar.HOUR_OF_DAY)==0){
\end{lstlisting}
\end{comment}

%<@\note{Source: org/apache/commons/math/analysis/solvers/BracketingNthOrderBrentSolver}@>
\begin{figure}
\begin{lstlisting}[style=java]
<@\note{Class: org.apache.commons.math.analysis.solvers.BracketingNthOrderBrentSolver}@>
<@\note{Method:protected double doSolve()}@>
<@\textbf{Developer patch:}@>
233: if (agingA >= MAXIMAL_AGING) {
234: // ...
235: <@\textcolor{red}{\textbf{-}}@>  targetY = -REDUCTION_FACTOR * yB;
236: <@\textcolor{green}{\textbf{+}}@>  final int p = agingA - MAXIMAL_AGING;
237: <@\textcolor{green}{\textbf{+}}@>  final double weightA = (1 << p) - 1;
238: <@\textcolor{green}{\textbf{+}}@>  final double weightB = p + 1;
239: <@\textcolor{green}{\textbf{+}}@>  targetY = (weightA * yA - weightB * REDUCTION_FACTOR * yB) / (weightA + weightB);
240:    } else if (agingB >= MAXIMAL_AGING) {
241: <@\textcolor{red}{\textbf{-}}@>   targetY = -REDUCTION_FACTOR * yA;
243: // ...
243: <@\textcolor{green}{\textbf{+}}@>   final int p = agingB - MAXIMAL_AGING;
244: <@\textcolor{green}{\textbf{+}}@>   final double weightA = p + 1;
245: <@\textcolor{green}{\textbf{+}}@>   final double weightB = (1 << p) - 1;
246: <@\textcolor{green}{\textbf{+}}@>   targetY = (weightB * yB - weightA * REDUCTION_FACTOR * yA) / (weightA + weightB);
\end{lstlisting}
\begin{lstlisting}[style=java]
<@\textbf{Patch $\patch_4$, generated by PraPR:}@>
260: <@\textcolor{red}{\textbf{-}}@>   if (signChangeIndex - start >= end - signChangeIndex) {
260: <@\textcolor{green}{\textbf{+}}@>   if (MAXMAL_AGING - start >= end - signChangeIndex) {
261:        ++start;
262:     } else {
263:        --end;
264:     }
\end{lstlisting}
\begin{lstlisting}[style=java]
<@\textbf{Patch $\patch_5$, generated by PraPR:}@>
317: <@\textcolor{red}{\textbf{-}}@>   x[signChangeIndex] = nextX;
317: <@\textcolor{green}{\textbf{+}}@>   x[agingA] = nextX;
318:     System.arraycopy(y, signChangeIndex, y, signChangeIndex + 1, nbPoints - signChangeIndex);
319:     y[signChangeIndex] = nextY;
\end{lstlisting}
\caption{\label{fig:math40} Developer and generated patches for Math-40}
%<@\rule{\columnwidth}{0.5pt}@>
\end{figure}

%<@\note{Source: com/google/javascript/jscomp/NodeUtil}@>
\begin{figure}
\begin{lstlisting}[style=java]
<@\note{Class: com.google.javascript.jscomp.NodeUtil}@>
<@\note{Method:static boolean functionCallHasSideEffects}@>
<@\textbf{Developer patch:}@>
958: <@\textcolor{green}{\textbf{+}}@>   if (nameNode.getFirstChild().getType() == Token.NAME) {
959: <@\textcolor{green}{\textbf{+}}@>   String namespaceName = nameNode.getFirstChild().getString();
960: <@\textcolor{green}{\textbf{+}}@>   if (namespaceName.equals("Math")) {
961: <@\textcolor{green}{\textbf{+}}@>       return false;
962: <@\textcolor{green}{\textbf{+}}@>       }
963: <@\textcolor{green}{\textbf{+}}@>   }
\end{lstlisting}

\begin{lstlisting}[style=java]
<@\textbf{Patch $\patch_{10}$, generated by PraPR:}@>    
933: <@\textcolor{red}{\textbf{-}}@>   if (callNode.isNoSideEffectsCall()) {
933: <@\textcolor{green}{\textbf{+}}@>   if (callNode.hasChildren()) {
934:        return false;
935:    }
\end{lstlisting}

\caption{\label{fig:closure61} Developer and generated patches for Closure-61}
%<@\rule{\columnwidth}{0.5pt}@>
\end{figure}

\begin{table}

\caption{Five top-ranked methods from Math-40}\label{table:bug1}
\begin{adjustbox}{width=1\columnwidth}
            \begin{tabular}{c|l|c|c|c|c}
            \hhline
            
       \textbf{EID} & \multicolumn{1}{c|}{\textbf{Method Signature}} & \textbf{\sbfl{}} & \textbf{PID} & \textbf{\#F} (1) & \textbf{\#P} (3177) \\ \hline
        
      $\meth_1$  &  \texttt{incrementEvaluationCount()} &0.57 & $\patch_1$ & 1 & 3170 \\ \hline
    $\meth_2$  & \texttt{BracketingNthOrderBrentSolver(Number)} & 0.33 & $\patch_2$ & 1 & 3172 \\ \hline
       $\meth_3$  & \texttt{BracketingNthOrderBrentSolver(double, ...)} & 0.28 & $\patch_3$ & 1 & 3177 \\ \hline

     \cellcolor{GrayOne}& \cellcolor{GrayOne}  & \cellcolor{GrayOne}  & \cellcolor{GrayOne}$\patch_4$ &\cellcolor{GrayOne}0 &\cellcolor{GrayOne} 3177 \\
    \cline{4-5}
    \multirow{-2}{*}{\cellcolor{GrayOne}$\meth_4^\mathbf{*}$}&\multirow{-2}{*}{\cellcolor{GrayOne}\texttt{doSolve()}}&\multirow{-2}{*}{\cellcolor{GrayOne}0.27}& \cellcolor{GrayOne}$\cellcolor{GrayOne}\patch_5$ & \cellcolor{GrayOne}0 & \cellcolor{GrayOne}3169\\ \hline  
    $\meth_5$  & \texttt{guessX(double[], ...)} & 0.20 &$\patch_6$ & 0 & 3176 \\ \hline
        \end{tabular}
        \end{adjustbox}
\end{table}

\begin{table}
    \caption{Five top-ranked methods from Closure-61}\label{table:bug2}
\begin{adjustbox}{width=1\columnwidth}
            \begin{tabular}{c|l|c|c|c|c}
            \hhline
            
        \textbf{EID} & \multicolumn{1}{c|}{\textbf{Method Signature}} & \textbf{SBFL} & \textbf{PID} & \textbf{\#F} (3) & \textbf{\#P} (7082) \\ \hline
        
        $\meth_1$ & \texttt{toString()} & 0.34 &$\patch_7$ & 3 & 7079 \\ \hline
        $\meth_2$ & \texttt{getSortedPropTypes()} & 0.33 &$\patch_8$ & 3 & 6981 \\ \hline
        $\meth_3$ & \texttt{toString(StringBuilder, ...)} & 0.27 &$\patch_9$& 3 & 7042 \\ \hline
        \cellcolor{GrayOne}$\meth_4^\mathbf{*}$ &\cellcolor{GrayOne}\texttt{functionCallHasSideEffects(Node, ...)}&\cellcolor{GrayOne}0.18 &\cellcolor{GrayOne}$\patch_{10}$&\cellcolor{GrayOne} 1 &\cellcolor{GrayOne}6681 \\ \hline
        $\meth_5$ & \texttt{nodeTypeMayHaveSideEffects(Node, ...)}& 0.09 &$\patch_{11}$  & 1 & 6766 \\ \hline
        \end{tabular}
        \end{adjustbox}
\end{table}

\section{Motivation Examples}\label{sec:motivation}
In this section, we present two real-world bug examples to show the limitations of the widely used \sbfl{} \fl{} and also the potential benefits that we can obtain from program repair.

\subsection{Example 1: Math-40}

We use Math-40 from \olddefects{}~\cite{bib:JJE14}, a widely used collection of real-world Java bugs, as our first example. Math-40 denotes the 40th buggy version of Apache Commons Math project~\cite{bib:commonsMath} from \olddefects{}. The bug is located in a single method of the project (method \texttt{doSolve} of class \texttt{BracketingNthOrderBrentSolver}).

We attempted to improve the effectiveness of traditional \sbfl{} based on Ochiai formula~\cite{abreu2007accuracy}, which has been widely recognized as one of the most effective \sbfl{} formulae~\cite{zhang2017boosting, li2017transforming, pearson2017evaluating}. Inspired by prior work~\cite{sohn2017fluccs}, we used the aggregation strategy to aggregate the maximum suspiciousness values from statements to methods. Even with this improvement in place, Ochiai still cannot rank the buggy method in the top, and instead ranks the buggy method in the 4th place (with a suspiciousness value of 0.27). The reason is that traditional \sbfl{} captures only coverage information and does not consider the actual impacts of code elements on test behaviors.
%Math-40 refers to Apache Commons Math 3.3 with a bug from the widely-used bug database \olddefects{}.  which has one failed test and one bug method related to the bug.
\Comment{
We improve the \naive \sbfl{} with aggregation strategy to aggregate suspiciousness value from statement to method, which is shown to be more effective than the \naive \sbfl{}~\cite{sohn2017fluccs}. Specifically, we use the widely used Ochiai formula~\cite{abreu2007accuracy}. Ochiai ranks the buggy method with the fourth high suspicious score 0.27. It is a common unsuccessful case for \sbfl{}, since it takes coverage as input, which is not precise enough to capture the test behaviors.
}

In an attempt to fix the bug, we further applied state-of-the-art \apr{} technique, \prapr{}~\cite{DBLP:conf/issta/GhanbariBZ19}, on the bug. However, since fixing the bug requires multiple lines of asymmetric edits, the genuine patch is beyond the reach of \prapr{} and virtually other existing \apr{} techniques as well. Analyzing the generated patches and their execution results, however, gives some insights on the positive effects that an \apr{} technique might have on \fl{}.
\Comment{
We further apply the latest \apr{} technique Prapr~\cite{DBLP:conf/issta/GhanbariBZ19} to fix this bug, but this bug cannot be fixed since the genuine patch is rather complicated.
After manually checking the generated patches and their execution results, we find there is some hint for \fl{}.}

Among a large number of methods in Math-40, Table~\ref{table:bug1} lists the Top-5 most suspicious methods based on Ochiai. Each row corresponds to a method, with the highlighted one corresponding to the actual buggy method (i.e., \texttt{doSolve}). Column ``EID'' assigns an identifier for each method. Column ``SBFL'' reports \sbfl{} suspiciousness values for each method, and ``PID'' assigns an identifier for each patch generated by \prapr{} that targets the method. Columns ``\#F'' and ``\#P'' report the number of failing and passing tests on each generated patch, respectively. The numbers within the parentheses in the table head are the number of failing/passing tests on the original buggy program. We also present the details of the developer patch for the bug and two patches generated by \prapr{} on the buggy method in Figure~\ref{fig:math40}. From the table, we observe that $\patch_4$ is a plausible patch, meaning that it passes all of the available tests but it might be not a genuine fix; $\patch_5$ passes originally failing tests, while fails to pass 8 originally passing tests.
\Comment{
We present the information of the bug in Table~\ref{table:bug1}, each row represents the information of each suspicious element (i.e., method), here we present only the highly-ranked methods considering the large number of methods. Column ``EID'' is the identifier of each element, the one annotated with * is the bug, i.e., $\meth_4$; Column ``Name'' presents the method name; ``SBFL'' shows the suspiciousness value for each method; ``PID'' is the identifier of each patch; ``Fail''/ ``Pass'' is the number of failed/passed tests on each patch, and the number in the bracket is the number of failed and passed tests on the original buggy program.We also present the details of the developer patch on the bug and two patches generated by \prapr{} on the bug method in Figure~\ref{fig:math40}.
$\patch_4$ is a plausible patch, on which although all the tests pass but not a genuine fix.
$\patch_5$ is a partially-fixed patch, on which although the  failed test becomes passing but 8 passed tests now fail.}

Several observations can be made at this point:
First, whether the originally failing tests pass or not on a patch, can help distinguish the buggy methods from some correct methods\Comment{, as well as some of correct methods}. For example, for $\meth_1$, $\meth_2$ and $\meth_3$, the originally failing test remains failing on all of their patches, while for the buggy method $\meth_4$, the originally failing test becomes passing on both its patches.
Second, whether the originally passing tests fail or not, can also help separate the buggy methods from some correct methods, e.g., $\patch_4$ for the buggy method $\meth_4$ does not fail any originally passing tests while the patch for the correct method $\meth_5$ still fails some originally passing tests.
Lastly, the detailed number of tests affected by the patches may not matter much. For example, for the correct method $\meth_5$, its patch only fails one originally passing test, but for the buggy method $\meth_4$, patch $\patch_5$ makes even more (i.e., 8) originally passing tests fail.

\Comment{
We have several observations.
First, the criteria whether failed tests become passing, can separate the buggy methods and part of correct methods apart.
For example, for $\meth_1$, $\meth_2$ and $\meth_3$, the failed tests are still failed on their patches.
Second, the criteria whether passed tests become failed, can also separate the buggy and correct methods, such as 
 $\meth_4$  and $\meth_5$.
Third, the detailed number of tests affected does not really matter. 
For example, for the correct method  $\meth_5$, its patch only fails one passed test,  but for the buggy method  $\meth_4$, its second patch $\patch_2$ makes 8 passed test become failed.
}

\subsection{Example 2: Closure-61}
\Comment{In order to confirm that the observations above are not merely coincidence,We now present another example to further illustrate our motivation in other cases. In particular, }We further looked into Closure-61, another real-world buggy project from \olddefects{}, but for which \prapr{} is even \textit{unable} to generate any plausible patch. \Comment{We next present how the execution information of non-plausible fixes can still improve the \sbfl{} results for this case. } Similar with the first example, we present the Ochiai fault localization information and \prapr{} repair results for the Top-5 methods in Table~\ref{table:bug2}. 
\Comment{We further manually check more cases to confirm that,  the observations above are not coincident, and since the first example has one plausible patch that directly occurs at the buggy method, and helps to point out the buggy method easily. We further look into the cases where although \apr{} cannot generate plausible patch, the execution information of partial fix patches still can improve \sbfl{} results.}

Based on Table~\ref{table:bug2}, we observe that even the non-plausible noisy patch $\patch_{10}$ is related to the buggy methods. The patches targeting method \texttt{getSortedPropTypes} and the two overloading methods of \texttt{toString} (which have higher suspiciousness values than that of the buggy method \texttt{functionCallHasSideEffects}) cannot generate any patch that can pass any of the originally failing tests.
\Comment{From the Table~\ref{table:bug2}, it's interesting to find that, even the partial fix patches are often related to the buggy methods.
For the methods \texttt{toString}, \texttt{toStringV} and  \texttt{getSPTypes} that have higher suspiciousness score than the buggy method \texttt{HasSideEffects}, they cannot generate any patches that turn failed tests into passing status.
It further shows that the patch execution results with even partial fix, can still suppress interference from the methods over-ranked by \sbfl{}.}
In addition, the fact that the number of passed tests which now fail in the patches of the buggy method\Comment{ \texttt{functionCallHasSideEffects}} are much larger than that for the correct method \texttt{nodeTypeMayHaveSideEffects} further confirms our observation above that, the detailed impacted test number does not matter much with the judgement of the correctness of a method.

Based on the above two examples, we have following implications to utilize the patch execution results to improve the original \sbfl{}: (1) the patches (no matter plausible or not) positively impacting some failed test(s) may indicate the actual buggy locations and should be favored; (2) the patches negatively impacting some passed test(s) may help exclude some correct code locations and should be unfavored; (3) the detailed number of the impacted tests does not matter much for fault localization. Therefore, we categorize all the patches into four different basic groups based on whether they impact originally passed/failed tests to help with fault localization, details shown in Section~\ref{sec:ap}.

\section{Approach}\label{sec:ap}

\subsection{Preliminaries}\label{sec:def}

\Comment{For a given patch, based on its execution results that (i) whether failed tests become pass and (ii) whether pass tests become failed, it can be  categorized into one of the four \patchgroup{}s as shown in Table~\ref{table:group}.
The column "ID" is the short identifier for the name of each \patchgroup{}, and we use \fp{} to represent the number of the failed tests which turn into pass on the patch, and \pf{} to represent the number of pass tests which turn into failed on the patch.
The column "Description" explains each \patchgroup{} in detail.}

%In this section, we formally define the terms that will be used across our approach section:
In order to help the readers better understand the terms used throughout this paper, in what follows, we attempt to define a number of key notions more precisely.

\begin{definition}[Candidate Patch]
Given the original program $\prog$, a candidate patch $\patch$ can be created by modifying one or more program elements within $\prog$. The set of all candidate patches generated for the program is denoted by $\patches$.
\end{definition}

In this paper, we focus on the \apr techniques that conduct only \emph{first-order} program transformations, which only change one program element in each patch, such as \prapr~\cite{DBLP:conf/issta/GhanbariBZ19}. Note that our approach is general and can also be applied to other \apr techniques in theory, even including the ones applying \emph{high-order} program transformations. \Comment{We emphasize that even with this simplifications, these definitions would be general enough to take into account almost all of the state-of-the-art \apr techniques, including \prapr~\cite{DBLP:conf/issta/GhanbariBZ19}.}
%Note that although our approach can be generalized to any \apr{} techniques in theory, in this work, we mainly focus on \emph{first-order} \apr{} techniques (i.e., each patch only modifies one code element), such as \prapr~\cite{ghanbari2019practical}.\Comment{ We denote the patches on a specific program element $\meth$ as $\patchLoc{\meth}$.}

\begin{definition}[Patch Execution Matrix]
\label{def:patch}
Given a program $\prog$, its test suite $\tsts$, and its corresponding set of all candidate patches $\patches$, the patch execution matrix, $\pMatrix$, is defined as the execution results of all patches in $\patches$ on all tests in $\tsts$. Each matrix cell result, $\pMatrixCell{\patch}{\tst}$, represent the execution results of test $\tst\in\tsts$ on patch $\patch\in\patches$, and can have the following possible values, \{\pass, \fail, \unknown\}, which represent \textit{failed}, \textit{passed}, and \textit{unknown yet}.
\end{definition}

Note that for the ease of presentation, we also include the original program execution results in $\pMatrix$, i.e., $\pMatrixCell{\prog}{\tst}$ denotes the execution results of test $\tst$ on the original program $\prog$.

Based on the above definitions, we can now categorize candidate patches based on the insights obtained from motivating examples:
\begin{definition}[\patchoneL] 	\label{def:patch1}
A patch $\patch$ is called a \patchoneL{}, i.e., $\group{\patch}=\patchoneS$, if it passes some originally failing tests while does not fail any originally passing tests, i.e., $\exists \tst\in\tsts, \pMatrixCell{\prog}{\tst}=$\fail$\wedge\pMatrixCell{\patch}{\tst}=$\pass, and $\nexists \tst\in\tsts, \pMatrixCell{\prog}{\tst}=$\pass$\wedge\pMatrixCell{\patch}{\tst}=$\fail.
\end{definition}
Note that $\group{\patch}$ returns the category group for each patch $\patch$.

\begin{definition}[\patchtwoL] \label{def:patch2}
A patch $\patch$ is called a \patchtwoL{}, i.e., $\group{\patch}=\patchtwoS$, if it passes some originally failing tests but also fails on some originally passing tests, i.e., $\exists \tst\in\tsts, \pMatrixCell{\prog}{\tst}=$\fail$\wedge\pMatrixCell{\patch}{\tst}=$\pass, and $\exists \tst\in\tsts, \pMatrixCell{\prog}{\tst}=$\pass$\wedge\pMatrixCell{\patch}{\tst}=$\fail.
\end{definition}

\begin{definition}[\patchthreeL] \label{def:patch3}
A patch $\patch$ is called a \patchthreeL{}, i.e., $\group{\patch}=\patchthreeS$, if it does not impact any originally failing or passing tests. More precisely, $\nexists \tst\in\tsts, \pMatrixCell{\prog}{\tst}=$\fail$\wedge\pMatrixCell{\patch}{\tst}=$\pass, and $\nexists \tst\in\tsts, \pMatrixCell{\prog}{\tst}=$\pass$\wedge\pMatrixCell{\patch}{\tst}=$\fail.
\end{definition}

\begin{definition}[\patchfourL] \label{def:patch4}
A patch $\patch$ is called a \patchfourL{}, i.e., $\group{\patch}=\patchfourS$, if it does not pass any originally failing test while fails some originally passing tests, i.e., $\nexists \tst\in\tsts, \pMatrixCell{\prog}{\tst}=$\fail$\wedge\pMatrixCell{\patch}{\tst}=$\pass\Comment{$\forall \tst\in\tsts, \pMatrixCell{\prog}{\tst}=$\fail$\implies\pMatrixCell{\patch}{\tst}=$\pass}, and $\exists \tst\in\tsts, \pMatrixCell{\prog}{\tst}=$\pass$\wedge\pMatrixCell{\patch}{\tst}=$\fail.
\end{definition}
%\dan{please double check the correctness of this formalization.}\lingming{I've corrected the original and Ali's modification}

Based on our insights obtained from the motivating example, the ranking of different patch groups is: $\patchoneS$ > $\patchtwoS$ > $\patchthreeS$ > $\patchfourS$. Note that in Section~\ref{sec:variant}, we will discuss more patch categorization variants besides such default patch categorization to further study their impacts on \ourtool{}.

\begin{figure}
    \centering
    \includegraphics[width=9cm, height= 4cm]{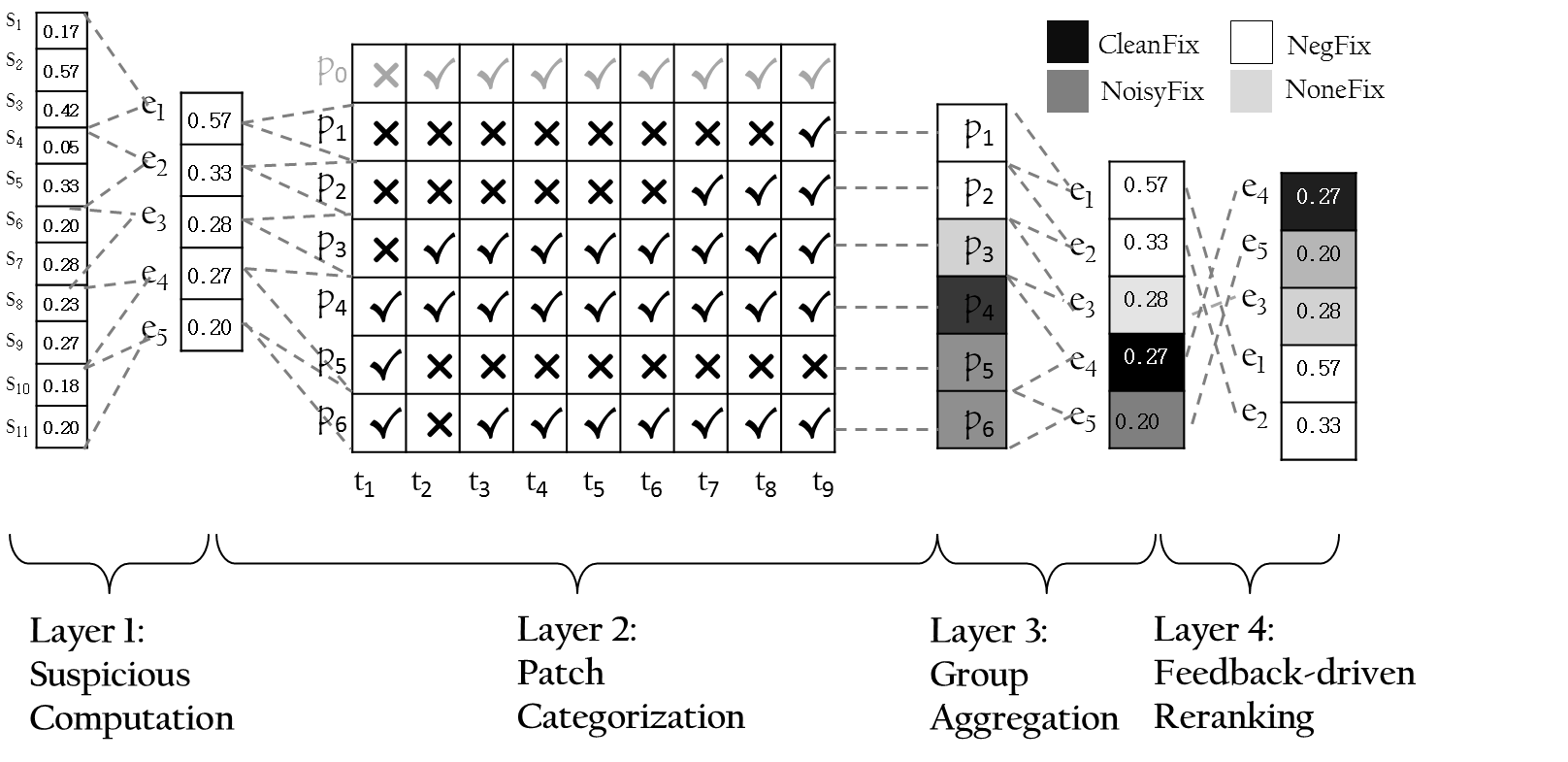}
    \caption{Overview of \ourtool{}}
    \label{fig:overview}
\end{figure}

\subsection{Basic \ourtool{}}\label{sec:basic}
The overview of \ourtool{} is shown in Figure~\ref{fig:overview}. According to the figure, \ourtool{} consists of four different layers. The input for \ourtool{} is the actual buggy program under test and the original failing test suite, and the final output is a refined ranking of the program elements based on the initial suspiciousness calculation.
In the first layer, \ourtool{} conducts naive \sbfl{} formulae (e.g., Ochiai~\cite{abreu2007accuracy}) at the statement level, and then perform \emph{suspiciousness aggregation}~\cite{sohn2017fluccs} to calculate the initial suspiciousness value for each program element. Note that besides such default initial suspiciousness computation, \ourtool{} is generic and can leverage the suspiciousness values computed by any other advanced fault localization technique in this layer (such as the PageRank-based fault localization~\cite{zhang2017boosting}). In the second layer, \ourtool{} collects the patch execution matrix along the program repair process for the program under test, and categorizes each patch into different groups based on Section~\ref{sec:def}. In the third layer, for each element, \ourtool{} maps the group information of its corresponding patches to itself via \emph{group aggregation}. In the last layer, \ourtool{} finally reranks all the program elements via considering their suspiciousness and group information in tandem.

We next explain each layer in details with our first motivation example. Since the number of tests and patches are really huge, due to space limitation, we only include the tests and patches that are essential for the ranking results of the elements.
After reduction, we consider the six patches ($\patch_1$ to $\patch_6$) and the 9 tests whose statuses changed on these patches (denoted as \tst$_1$ to  \tst$_9$).
Based on Definition~\ref{def:patch}, we present $\pMatrix$ in Figure~\ref{fig:overview}, The first row stands for $\pMatrixCell{\prog}{\tsts}$, the execution results of $\tsts$ on the original buggy program $\prog$ , and from the second row, each row represents $\pMatrixCell{\patch}{\tsts}$, the execution results of each patch $\patch$ as shown in Table~\ref{table:bug1} on $\tsts$.

 \begin{comment}

 \begin{equation}\label{eq:matrix}
     \pMatrix = \begin{bmatrix} 
    & \prog & \patch_1 & \patch_2 & \patch_3 & \patch_4 & \patch_5 & \patch_6 \\
     \tst_1 & $\fail$ & $\fail$ & $\fail$ & $\fail$ & $\pass$ &  $\pass$ & $\pass$ \\
    \tst_2 & $\pass$ & $\fail$ &  $\fail$ & $\pass$ & $\pass$ & $\fail$ & $\fail$ \\
     \tst_3 & $\pass$ & $\fail$ & $\fail$ & $\pass$ & $\pass$ & $\fail$ & $\pass$ \\
     \tst_4 & $\pass$& $\fail$ & $\fail$ &$\pass$& $\pass$& $\fail$ &$\pass$ \\
   \tst_5 & $\pass$&$\fail$&$\fail$&$\pass$& $\pass$& $\fail$& $\pass$ \\
    \tst_6 & $\pass$&$\fail$& $\fail$&$\pass$&$\pass$& $\fail$& $\pass$ \\
    \tst_7 & $\pass$& $\fail$& $\pass$& $\pass$& $\pass$& $\fail$& $\pass$\\
     \tst_8 & $\pass$& $\fail$& $\pass$& $\pass$& $\pass$& $\fail$ & $\pass$ \\
     \tst_9 & $\pass$& $\pass$ & $\pass$& $\pass$& $\pass$&$\fail$ & $\pass$ \\
     \end{bmatrix}\\
     \end{equation} 
\end{comment}

\subsubsection{Layer 1: Suspicious Computation} \label{sec:lay1}
Given the original program statements, e.g., $\stmts=[\stmt_1, \stmt_2,...,\stmt_n]$,  we directly apply an off-the-shelf spectrum-based fault localization technique (e.g., the default Ochiai~\cite{abreu2007accuracy}) to compute the suspiciousness for each statement, e.g., $\susp{\stmt_j}$ for statement $\stmt_j$. Then, the proposed 
approach applies \emph{suspiciousness aggregation}~\cite{sohn2017fluccs} to compute the element suspiciousness values at the desired level (e.g., method level in this work) since prior work has shown that suspicious aggregation can significantly improve fault localization results~\cite{sohn2017fluccs, chen2019compiler}. 
Given the initial list of $\meths=[\meth_1, \meth_2,...,\meth_m]$, for each $\meth_i\in \meths$, suspiciousness aggregation computes its suspiciousness as $\susp{\meth_i}=Max_{\stmt_j\in\meth_i} \susp{\stmt_j}$, i.e., the highest suspiciousness value for all statements within a program element is computed as the suspiciousness value for the element.

For our first motivation example, after suspicious aggregation, for the five elements, 
$\susp{\meth_1,\meth_2,\meth_3,\meth_4,\meth_5} = [0.57, 0.33, 0.28, 0.27, 0.20]$.

%\lingming{add the example illustration to each step, e.g., showing the actual patch execution matrix for the example in motivation}
%\lingming{also, try to make these 4 layers look more complicated, I tried but my eyes hurt}

\subsubsection{Layer 2: Patch Categorization} \label{sec:lay2}
In this layer, \ourtool{} automatically invokes off-the-shelf program repair engines (\prapr{}~\cite{DBLP:conf/issta/GhanbariBZ19} for this work) to try various patching opportunities and record the detailed patch-execution matrix, $\pMatrix$. Then, based on the resulting $\pMatrix$, \ourtool{} automatically categorizes each patch into different groups. Given program element $\meth$ and all the patches generated for the program, $\patches$, the patches occurring on $\meth$ can be denoted as $\patchLoc{\meth}$. Then, based on Definitions~\ref{def:patch1} to \ref{def:patch4}, each patch within $\patchLoc{\meth}$ for each element $\meth$ can be categorized into one of the four following groups, \{$\patchoneS$, $\patchtwoS$, $\patchthreeS$, $\patchfourS$\}. Recall that $\group{\patch}$ represents the group information for $\patch$, e.g.,
$\group{\patch}=\patchoneS$ denotes that $\patch$ is a clean-fix patch.

For the example, the group of each patch in the motivation example is as follows:
$\group{\patch_1,\patch_2, \patch_3, \patch_4, \patch_5, \patch_6} =  [\patchfourS, \patchfourS,$ $\patchthreeS,\patchoneS, \patchtwoS, \patchtwoS] $

\subsubsection{Layer 3: Group Aggregation}\label{sec:lay3}

For each program element $\meth$, we utilize its corresponding patch group information to determine its own group information. Recall that the ranking of different patch groups is: $\patchoneS$>$\patchtwoS$>$\patchthreeS$>$\patchfourS$. Then, the group information for a program element can be determined by the best group information of all patches occurring on the program element. Therefore, we present the following rules for determining the group information for each $\meth$:
\vspace*{-5pt}
\begin{equation}\label{eq:cagg}
    \group{\meth} = 
    \begin{cases}
    \patchoneS{} & \text{if $\exists \patch,~ \patch\in\patchLoc{\meth}\wedge\group{\patch}=\patchoneS$}  \\
    \patchtwoS{} & \text{else if $\exists \patch,~ \patch\in\patchLoc{\meth}\wedge\group{\patch}=\patchtwoS$}  \\
    \patchthreeS{} & \text{else if $\exists \patch,~ \patch\in\patchLoc{\meth}\wedge\group{\patch}=\patchthreeS$}  \\
    \patchfourS{} & \text{else if $\exists \patch,~ \patch\in\patchLoc{\meth}\wedge\group{\patch}=\patchfourS$} 
    \end{cases}
\end{equation}

Shown in Equation~\ref{eq:cagg}, element $\meth$ is within Group \patchoneS{} whenever there is any patch $\patch$ within $\meth$ such that $\patch$ is a clean-fix patch; otherwise, it is within Group \patchtwoS{} whenever there is any patch $\patch$ within $\meth$ such that $\patch$ is a noisy-fix patch.

After group aggregation, the group of each program element (i.e., method) in the motivation example is  
$\group{\meth_1,\meth_2, \meth_3, \meth_4, \meth_5} = [\patchfourS, \patchfourS,$ $\patchthreeS,\patchoneS, \patchtwoS]$.

\subsubsection{Layer 4: Feedback-driven Reranking} \label{sec:lay4}

In this last layer, we compute the final ranked list of elements based on the aggregated suspiciousness values and groups. All the program elements will be first clustered into different groups with Group \patchoneS{} ranked first and Group \patchfourS{} ranked last. Then, within each group, the initial \sbfl{} (or other  fault localization techniques) suspiciousness values will be used to rank the program elements. \Comment{Assume we use $\rank$ to denote the total-order ranking between all program elements based on both the aggregated suspiciousness and groups} Assume we use $\prelation{\meth_1, \meth_2}$ to denote the total-order ranking between any two program elements, \Comment{ to calculate the ranking between $\meth_1$ and $\meth_2$}it can be formally defined as:
\vspace*{-5pt}
\begin{equation}\label{eq:rerank}
    \prelation{\meth_1, \meth_2} = 
    \begin{cases}
     \meth_1 \rank \meth_2 &\text{if $\group{\meth_1}>\group{\meth_2}$ or }  \\ 
     &\text{ $\group{\meth_1}=\group{\meth_2}\wedge \susp{\meth_1}\geq\susp{\meth_2}$ }  \\

    \meth_2 \rank \meth_1 &\text{if $\group{\meth_2}>\group{\meth_1}$ or }  \\ 
     &\text{ $\group{\meth_1}=\group{\meth_2}\wedge \susp{\meth_2}\geq\susp{\meth_21}$ }  
    \end{cases}
\end{equation}

That is, $\meth_1$ is ranked higher or equivalent to $\meth_2$ only when (i) $\meth_1$ is within a higher-ranked group, or (ii) $\meth_1$ is within the same group as $\meth_2$ but has a higher or equivalent suspicious value compared to $\meth_2$.
Therefore, the final ranking of our example is: $\meth_4 \rank \meth_5  \rank  \meth_3 \rank \meth_1 \rank \meth_2$, ranking the buggy method $\meth_4$ at the first place.

\subsection{Variants of \ourtool{}}\label{sec:variant}
Taking the approach above as the basic version of \ourtool{}, there can be many variants of \ourtool{}, which are discussed as follows.

\begin{figure}
    \centering
    \includegraphics[width=8.3cm, height=2.8cm]{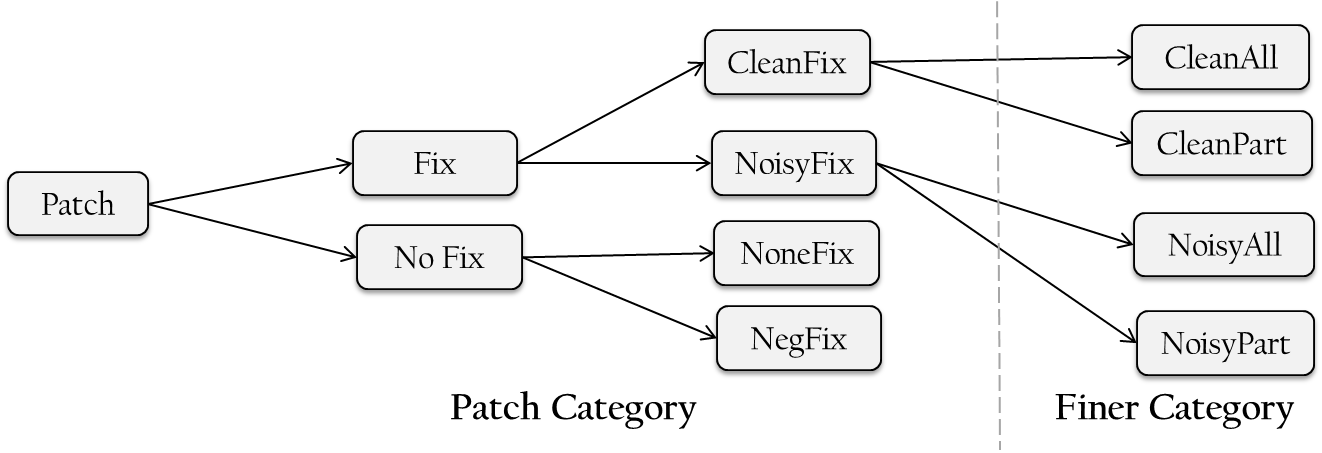}
            \caption{Patch categorization tree}
            \label{fig:tree}
\end{figure}

\parabf{Finer-grained Patch Categorization.}
Previous work~\cite{DBLP:conf/issta/GhanbariBZ19} found that plausible patches are often coupled tightly with buggy elements, which actually is a subset of  \patchoneS{} defined in our work.
Inspired by this finding, we further extend \ourtool{} with finer-grained patch categorization rules, which respectively divide \patchoneS{} and \patchtwoS{} into two finer categories according to the criterion whether all failed tests are impacted.
We use Figure~\ref{fig:tree} to show the relation between the four finer-grained patch categories and the four basic categories.
Considering the finer categories, we further extend the group aggregation strategies in the third layer of \ourtool{} accordingly as shown in Table~\ref{table:variants} to study the impact of further splitting \patchoneS{} and \patchtwoS{} categories, e.g., $R_1$ and $R_2$ study the two different rules splitting \patchoneS{}.

\parabf{SBFL Formulae.} 
The elements are reranked in the last layer based on their 
aggregated suspiciousness values and groups.
In theory, \ourtool{} is not specific for any particular way  to calculate the aggregated suspiciousness value.
Therefore, besides our default Ochiai~\cite{abreu2007accuracy} formula, all the other formulae in \sbfl{} can be adopted in \ourtool{}\Comment{, which are presented in the Table~\ref{table:sbfl}\lingming{it should be table 4}}.
We study all the \numsbfl{} \sbfl{} formulae considered in prior work~\cite{sohn2017fluccs,li2017transforming}.
The impact of these formulae on \ourtool{} would be studied later.

\parabf{Feedback Sources.} 
Generally speaking, not only the patch execution results can be the feedback of our approach, any other execution results correlated with program modifications can serve as the feedback sources, e.g., mutation testing~\cite{DBLP:journals/tse/JiaH11}. For example, a mutant and a patch are both modifications on the program, thus \ourtool{} can directly be applied with the mutation information as feedback.
However, mutation testing often includes simple syntax modifications that were originally proposed to simulate software bugs to fail more tests, while program repair often includes more (advanced) modifications that aim to pass more tests to fix software bugs.
Therefore, although it is feasible to use mutation information as the feedback source of our approach, the effectiveness remains unknown, which would be studied.

\parabf{Partial Execution Matrix.} 
During program repair, usually the execution for a patch would terminate as soon as one test fails, which is the common practice to save the time cost.
In this scenario, only partial execution results are accessible. In the previous sections, $\pMatrix$ is considered as complete, which we denote as \textit{full} matrix, $\pMatrixfull$, while in this section, we discuss the case where $\pMatrix$ is considered as incomplete in practice, which we call a \textit{partial} matrix, $\pMatrixpart$.
Recall Definition~\ref{def:patch}, different from $\pMatrixfull$, the cells in $\pMatrixpart$ can be \unknown{}  besides \pass{} and \fail{}.
For example, when $\tst$ is not executed on $\patch$, $\pMatrixpartCell{\patch}{\tst}$ $=$ \unknown.

In the motivation example, during the patch execution, if $\tsts$ is executed in the order from $\tst_1$ to $\tst_9$, and one failed test would stop execution for each patch immediately, $\pMatrixpart$ is as follows.
\begin{equation}\label{eq:matrixp}\small
    \pMatrixpart = \begin{bmatrix} 
                & \tst_1 & \tst_2 & \tst_3 & \tst_4 & \tst_5 & \tst_6 & \tst_7& \tst_8 & \tst_9 \\ 
        \prog &  $\fail$ & $\pass$& $\pass$&$\pass$ &$\pass$ &$\pass$ &$\pass$&$\pass$ & $\pass$ \\
     \patch_1 &  $\fail$ & $\unknown$ & $\unknown$ & $\unknown$ & $\unknown$ & $\unknown$ & $\unknown$ & $\unknown$ & $\unknown$ \\    
     \patch_2 &  $\fail$ & $\unknown$ & $\unknown$ & $\unknown$ & $\unknown$ & $\unknown$ & $\unknown$ & $\unknown$ & $\unknown$  \\
     \patch_3 &  $\fail$ & $\unknown$ & $\unknown$ & $\unknown$ & $\unknown$ & $\unknown$ & $\unknown$ & $\unknown$ & $\unknown$  \\
     \patch_4 &  $\pass$ & $\pass$&$\pass$ &$\pass$ &$\pass$ &$\pass$&$\pass$ & $\pass$& $\pass$ \\
     \patch_5 &  $\pass$ & $\fail$ & $\unknown$ & $\unknown$ & $\unknown$ & $\unknown$ & $\unknown$ & $\unknown$ & $\unknown$ \\
     \patch_6 &  $\pass$ & $\fail$ & $\unknown$ & $\unknown$ & $\unknown$ & $\unknown$ & $\unknown$ & $\unknown$ & $\unknown$\\
    \end{bmatrix}\\
\end{equation}

\begin{comment}
\begin{equation}\label{eq:matrixp}
    \pMatrixpart = \begin{bmatrix} 
     & \prog & \patch_1 & \patch_2 & \patch_3 & \patch_4 & \patch_5 & \patch_6 \\
    \tst_1 & $\fail$ & $\fail$ & $\fail$ & $\fail$ & $\pass$ &  $\pass$ & $\pass$ \\
    \tst_2 & $\pass$ & $\unknown$ &  $\unknown$ & $\unknown$ & $\pass$ & $\fail$ & $\fail$ \\
    \tst_3 & $\pass$ & $\unknown$ & $\unknown$ & $\unknown$ & $\pass$ & $\unknown$ & $\unknown$ \\
  \tst_4 & $\pass$ & $\unknown$ & $\unknown$ &$\unknown$& $\pass$& $\unknown$ &$\unknown$ \\
    \tst_5 & $\pass$ & $\unknown$& $\unknown$&$\unknown$& $\pass$& $\unknown$&$\unknown$ \\
    \tst_6 & $\pass$ & $\unknown$& $\unknown$&$\unknown$&$\pass$& $\unknown$&$\unknown$ \\
    \tst_7 & $\pass$ & $\unknown$& $\unknown$& $\unknown$& $\pass$& $\unknown$&$\unknown$\\
    \tst_8 & $\pass$ & $\unknown$& $\unknown$& $\unknown$& $\pass$& $\unknown$ & $\unknown$ \\
    \tst_9 & $\pass$ &$\unknown$& $\unknown$& $\unknown$& $\pass$&$\unknown$ & $\unknown$ \\
    \end{bmatrix}\\
\end{equation} 
\end{comment}

\begin{table}
    \centering
        \caption{Finer-grained patch categorization rules}\label{table:variants}
\begin{adjustbox}{width=0.9\columnwidth}
\begin{tabular}{l|l}    
    \hhline
    \textbf{ID} & \textbf{Extended Categorization Aggregation Rules} \\ \hline
    
     $R_1$ &  $\smallpatchoneS$>$\smallpatchtwoS$>$\patchtwoS$>$\patchthreeS$>$\patchfourS$ \\ \hline
     
    $R_2$ &  $\smallpatchtwoS$>$\smallpatchoneS$>$\patchtwoS$>$\patchthreeS$>$\patchfourS$\\ \hline
    
    $R_3$& $\patchoneS$>$\smallpatchthreeS$>$\smallpatchfourS$>$\patchthreeS$>$\patchfourS$ \\ \hline
    
    $R_4$ & $\patchoneS$>$\smallpatchfourS$> $\smallpatchthreeS$>$\patchthreeS$>$\patchfourS$ \\
    
   \hhline
    \end{tabular}
    \end{adjustbox}
\end{table}

\begin{comment}

 \begin{equation}\label{eq:matrixp}
    \pMatrixpart = \begin{bmatrix} 
    
     & $\prog$ & $\patch_1$ & $\patch_2$ & $\patch_3$ & $\patch_4$ & $\patch_5$ & $\patch_6$ \\
    $\tst_1$ & $\fail$ & $\fail$ & $\fail$ & $\fail$ & $\pass$ &  $\pass$ & $\pass$ \\
     $\tst_2$ & $\pass$ & $\unknown$ &  $\unknown$ & $\unknown$ & $\pass$ & $\fail$ & $\fail$ \\
     $\tst_3$ & $\pass$ & $\unknown$ & $\unknown$ & $\unknown$ & $\pass$ & $\unknown$ & $\unknown$ \\
   $\tst_4$ & $\pass$ & $\unknown$ & $\unknown$ &$\unknown$& $\pass$& $\unknown$ &$\unknown$ \\
     $\tst_5$ & $\pass$ & $\unknown$& $\unknown$&$\unknown$& $\pass$& $\unknown$&$\unknown$ \\
    $\tst_6$ & $\pass$ & $\unknown$& $\unknown$&$\unknown$&$\pass$& $\unknown$&$\unknown$ \\
    $\tst_7$ & $\pass$ & $\unknown$& $\unknown$& $\unknown$& $\pass$& $\unknown$&$\unknown$\\
    $\tst_8$ & $\pass$ & $\unknown$& $\unknown$& $\unknown$& $\pass$& $\unknown$ & $\unknown$ \\
    $\tst_9$ & $\pass$ &$\unknown$& $\unknown$& $\unknown$& $\pass$&$\unknown$ & $\unknown$ \\
     \end{bmatrix}\\
 \end{equation} 

\end{comment}

In the scenario where only partial matrix is accessible, we can find there are many unknown results. 
\Comment{Even we can further fill in  the matrix with coverage information by keeping the tests which do not cover the patch with it original status \dan{Cannot understand this sentence.} (i.e., the same as $\pMatrixpartCell{\prog}{\tsts}$)~\cite{DBLP:conf/issta/GhanbariBZ19}, there still remains many unknown cells in the $\pMatrixpart$.}
Interestingly, in this example, we find the final ranking does not change at all even with a partial matrix as input.
For the patches $\patch_3$, $\patch_4$, $\patch_5$ and $\patch_6$, their patch categorization does not change at all. 
For example, since the failed tests are executed first, when $\patch_5$ stops its execution, its execution result is that one failed test passes now and one passed test fails now, and thus $\patch_5$ is still categorized into $\patchtwoS$.
For $\patch_1$ and $\patch_2$, although their patch categorization changes from \patchfourS{} to \patchthreeS{}, it does not impact the final ranking results.
The example indicates the insensitivity of \ourtool{} to partial matrix, and the categorization design is the main reason for it.
We would further confirm this observation in the detailed experimental studies.

\section{Experiment Set Up} \label{sec:setup}

\subsection{Research Questions}\label{sec:rq}
In our study, we investigate the following research questions:
\begin{itemize}
    \item \textbf{RQ1:} 
    How does the basic \ourtool{} perform compared with state-of-the-art \sbfl{} and \mbfl{} techniques?

    \item \textbf{RQ2:} 
    How do different experimental configurations impact \ourtool{}?

    \begin{itemize}
         \item \textbf{RQ2a:} What is the impact of finer patch categorization?
        \item \textbf{RQ2b:} What is the impact of the used \sbfl{} formula?
        \item \textbf{RQ2c:} What is the impact of the feedback source used?
         \item \textbf{RQ2d:} What is the impact of partial execution matrix?
         \item \textbf{RQ2e:} What is the impact of the used benchmark suite?
    \end{itemize}
    
    \item \textbf{RQ3:}
    Can \ourtool{} further boost state-of-the-art unsupervised- and supervised-learning-based \fl{}?

\end{itemize}

\Comment{RQ1 compares \ourtool{} with state-of-the-art \fl{} techniques to investigate the effectiveness of \ourtool{}. Note that we focus on state-of-the-art \sbfl{} and \mbfl{} fault localization techniques in this RQ since they are closely related. \Comment{since they require extensive training data which may not always be available.}
RQ2 is a series of research questions to evaluate the impact of different experimental configurations on \ourtool{}.
Lastly, RQ3 further studies whether \ourtool{} can further boost state-of-the-art learning-based \fl{}.}

\begin{table}
	\centering
	\small
	\caption{Benchmark statistics}\label{table:subinfo}
	\begin{tabular}{l l|l|l|l}
		\hhline
        \textbf{ID} & \textbf{Name}&\textbf{\#Bug} & \textbf{\#Test} &\textbf{LoC}\\ \hhline
	    Lang & Apache commons-lang & 65 & 2,245 & 22K \\ 
	    Math & Apache commons-math & 106 & 3,602 & 85K \\ 
	    Time & Joda-Time & 27 & 4,130 & 28K \\ 
	    Chart & JFreeChart & 26 & 2,205 & 96K \\ 
	    Closure & Google Closure compiler & 106 & 7,927 & 90K \\ 
	    Mockito & Mockito framework & 38 & 1,366 & 23K \\ \hline
	   \multicolumn{2}{c|}{\textbf{\olddefects{}}} & \textbf{395} & \textbf{21,475} & \textbf{344K}\\
		\hline
		\hline
		Cli & Apcache commons-cli & 24 & 409 & 4K \\ 
		Codec & Apache commons-codec & 22 & 883 & 10K \\
		Csv & Apache commons-csv & 12 & 319 & 2K \\
		JXPath & Apache commons-jxpath & 14 & 411 & 21K \\
		\cellcolor{GrayOne}Gson & \cellcolor{GrayOne}Google GSON & \cellcolor{GrayOne}16 & \cellcolor{GrayOne}N/A & \cellcolor{GrayOne}12K \\
		\cellcolor{GrayOne}Guava & \cellcolor{GrayOne}Google Guava & \cellcolor{GrayOne}9 & \cellcolor{GrayOne}1,701,947 & \cellcolor{GrayOne}420K \\
		Core & Jackson JSON processor & 13 & 867 & 31K \\
		Databind & Jackson data bindings & 39 & 1,742 & 71K \\
		Xml & Jackson XML extensions & 5 & 177 & 6K \\
		Jsoup & Jsoup HTML parser & 63 & 681 & 14K \\ \hline
		\multicolumn{2}{c|}{\textbf{\newdefects{}}} & \textbf{587} & \textbf{26,964} & \textbf{503K} \\ \hline
	\end{tabular}
	
\end{table}

\subsection{Benchmark} \label{sec:benchmark}
We conduct our study on all bugs from the Defects4J benchmark~\cite{bib:JJE14}, which has been widely used in prior fault-localization work~\cite{li2017transforming, sohn2017fluccs, li2019deepfl, zhang2017boosting, pearson2017evaluating}. 
Defects4J is a collection of reproducible real bugs with a supporting infrastructure.
To our knowledge, all the \fl{} studies evaluated on Defects4J use the original version \olddefects{}. Recently, an extended version, \newdefects{}, which includes more real-world bugs, has been released~\cite{newdefects4j}. 
\Comment{Therefore, we further study the impact from the \newdefects{} to reduce the threats to external validity.}
Therefore, we further perform the first fault localization study on \newdefects{} to reduce the threats to external validity.

We present the  details of the used benchmarks in Table~\ref{table:subinfo}. Column ``ID''  presents the subject IDs used in this paper. Columns ``Name'' and  ``\#Bugs'' present the full name and the number of bugs for each project. Columns ``Loc'' and ``\#Test'' list the line-of-code information and the number of tests for the HEAD version of each project. Note that the two projects highlighted in gray are excluded from our evaluation due to build/test framework incompatibility with \prapr{}~\cite{DBLP:conf/issta/GhanbariBZ19}. In total, our study is performed on all 395 bugs from \olddefects{} and 192 additional bugs from \newdefects{}.

\subsection{Independent Variables}

\parabf{Evaluated Techniques:}
We compare \ourtool{} with the following state-of-the-art \sbfl{} and \mbfl{} techniques:
\textbf{(a) Spectrum-based (\sbfl{})}: we consider traditional SBFL with suspiciousness aggregation strategy to aggregate suspiciousness values from statements to methods, which has been shown to be more effective than naive \sbfl{} in previous work~\cite{sohn2017fluccs,chen2019compiler}.
\textbf{(b) Mutation-based (\mbfl{})}: we consider two representative \mbfl{} techniques, \muse{}~\cite{moon2014ask} and \meta{}~\cite{papadakis2015metallaxis}. 
\textbf{(c) Hybrid of \sbfl{} and \mbfl{} (\hyb{})}: we also consider the recent \hyb{}~\cite{pearson2017evaluating}, which represents state-of-the-art hybrid spectrum- and mutation-based fault localization\Comment{ the suspiciousness score of \sbfl{} and \mbfl{} via average}. Furthermore, we also include state-of-the-art learning-based fault localization techniques: \textbf{(a) Unsupervised}: we consider state-of-the-art \pkbasic{}~\cite{zhang2017boosting} and \pkag{}~\cite{zhang2019empirical} (which further improves \pkbasic{} via suspiciousness aggregation) that aim to boost \sbfl{} with the unsupervised PageRank algorithm. \textbf{(b) Supervised}: we further consider state-of-the-art supervised-learning-based fault localization, \dpfl~\cite{li2019deepfl}, which outperforms all other learning-based fault localization~\cite{li2017transforming, sohn2017fluccs, xuan2014learning}. Note that, \sbfl{} and \meta{} can adopt different \sbfl{} formulae, and we by default uniformly use Ochiai~\cite{abreu2007accuracy} since it has been demonstrated to perform the best for both \sbfl{} and \mbfl~\cite{pearson2017evaluating, li2017transforming, zhang2017boosting}.

\parabf{Experimental Configurations:} We explore the following configurations to study \ourtool{}:
\textbf{(a) Finer \ourtool{} Categorization}: in RQ2a, we study the four extended categorization aggregation rules based on the finer patch categories as listed in Table~\ref{table:variants}. \textbf{(b) Studied \sbfl{} Formulae}: in RQ2b, we implement all the \numsbfl{} \sbfl{} formulae considered in prior work~\cite{sohn2017fluccs,li2017transforming} to study the impact of initial formulae.
%presented in Table~\ref{table:sbfl} \yiling{should we keep the table?}\lingming{no, you don't have space} to study the impact of initial formulae.
\textbf{(c) Feedback Sources}: besides the patch execution results of program repair, mutation testing results can also be used as the feedback sources of \ourtool{}. Thus, we study the impact of these two feedback sources in RQ2c. \textbf{(d) Partial Execution Matrix}:
we collect partial execution matrices in three common test-execution orderings: \textbf{(i) \executionorderone{}}: the default order in original test suite; \textbf{(ii) \executionordertwo{}}: running originally-failed tests first and then originally-passing tests, which is also the common practice in program repair to save the time cost;
\textbf{(iii) \executionorderthree}: running originally-passing tests first and then originally-failed tests.
The partial matrices collected by these orders are denoted as \orderone{}, \ordertwo{} and \orderthree{} respectively.
We investigate the impacts of different partial execution matrices used in RQ2d.
\textbf{(e) Used Benchmarks}: we evaluate \ourtool{} in two benchmarks, \olddefects{} and \newdefects{} in RQ2e.

\subsection{Dependent Variables and Metrics}

\Comment{In empirical studies, dependent variables (DV) are used to
indicate and measure the interesting aspects of final results}
In this work, we perform \fl{} at the method level following recent fault localization work~\cite{sohn2017fluccs, b2016learning, li2019deepfl, zhang2017boosting, li2017transforming}, because the class level has been shown to be too coarse-grained while the statement level is too fine-grained to keep useful context information~\cite{parnin2011automated, kochhar2016practitioners}. We use the following widely used metrics~\cite{li2017transforming, li2019deepfl}:

\parabf{Recall at \topn{}:} \topn{} computes the number of bugs with at least one buggy element localized in the Top-N positions of the ranked list. As suggested by prior work~\cite{parnin2011automated}, usually, programmers only inspect a few buggy elements in the top of the given ranked list, e.g., 73.58\% developers only inspect Top-5 elements~\cite{kochhar2016practitioners}.
Therefore, following prior work~\cite{zou2019empirical, li2017transforming, li2019deepfl, zhang2017boosting}, we use \topn{} (N=1, 3, 5).

%\dan{add some reference for MFR and MAR}
\parabf{Mean First Rank (MFR):} For each subject, MFR computes the mean of the first relevant buggy element's rank for all its bugs, because the localization of the first buggy element for each bug can be quite crucial for localizing all buggy elements.

\parabf{Mean Average Rank (MAR):} We first compute the average ranking of all the buggy elements for each bug. Then, MAR of each subject is the
mean of such average ranking of all its bugs. MAR emphasizes the precise ranking of all buggy elements, especially for the bugs with multiple buggy elements.

\Fl{} techniques sometimes assign same suspiciousness score to code elements. Following prior work~\cite{li2017transforming, li2019deepfl}, we use the \emph{worst} ranking for the tied elements. For example, if a buggy element is tied with a correct element in the $k^{th}$ position of the ranked list, the rank for both elements would be $k+1^{th}$.

\subsection{Implementation and Tool Supports}
For \apr{}, we use \PraPR{}~\cite{DBLP:conf/issta/GhanbariBZ19}, a recent \apr{} technique that fixes bugs at the bytecode level. 
We choose \PraPR{} because it is one of the most recent \apr{} techniques and has been demonstrated to be able to fix more bugs with a much lower overhead compared to other state-of-the-art techniques. 
Note that, \ourtool{} does not rely on any specific \apr{} technique, since the feedback input (i.e., patch-execution information) for our approach is general and can work with any other \apr{} technique in principle.

We now discuss the collection of all the other information for implementing \ourtool{} and other compared techniques: (i) To collect the coverage information required by \sbfl{} techniques, 
we use the ASM bytecode manipulation framework~\cite{bruneton2002asm} to instrument the code on-the-fly via JavaAgent~\cite{javaagent}.
(ii) To collect the mutation testing information required by \mbfl{}, we use state-of-the-art PIT mutation testing framework ~\cite{Pitest} (Version 1.3.2) with all its available mutators, following prior \mbfl{} work~\cite{li2017transforming, li2019deepfl}.
 Note that we also modify PIT to force it to execute all tests for each mutant and collect detailed mutant impact information (i.e., whether each mutant can impact the detailed test failure message of each test~\cite{pearson2017evaluating}) required by \meta{}. For \pkbasic{}, \pkag{}, and \dpfl, we directly used the implementation released by the authors~\cite{zhang2019empirical, li2019deepfl}.

All the experiments are conducted on a Dell workstation with Intel(R) Xeon(R) Gold 6138 CPU @ 2.00GHz and 330GB RAM, running Ubuntu 18.04.1 LTS.

\subsection{Threats to Validity}
\parabf{Threats to internal validity} mainly lie in the correctness of implementation of our approach and the compared techniques. 
To reduce this threat, we manually reviewed our code and verified that the results of the overlapping \fl{} techniques between this work and prior work~\cite{li2017transforming, zhang2017boosting,zhang2019empirical} are consistent. We also directly used the original implementations from prior work~\cite{zhang2019empirical, li2019deepfl}.

\parabf{Threats to construct validity} mainly lie in the rationality of assessment metrics that we chose. 
To reduce this threat, we chose the metrics that have been recommended by prior studies/surveys~\cite{parnin2011automated, kochhar2016practitioners} and widely used in previous work~\cite{li2017transforming,li2019deepfl,sohn2017fluccs,zhang2017boosting}.

\parabf{Threats to external validity} mainly lie in the benchmark suites used in our experiments. To reduce this threat, we chose the widely used Defects4J benchmark, which includes hundreds of real bugs collected during real-world software development.
To further reduce the threats, compared to previous work, we not only used the original version of Defects4J, but also conducted the first fault localization evaluation on an extended version of Defects4J.

\section{Results}

\begin{table}
    \centering
   \caption{Overall fault localization results}\label{table:cmp}
   	\begin{adjustbox}{width=0.75\columnwidth}
    \begin{tabular}{l|l|l|l|l|l}
    \hhline
     Tech Name & \topone{} & \topthree{} & \topfive{} & \mfr{} & \mar{} \\ \hline
     
 \sbfl{} & 117 & 219 & 259 & 19.15 & 24.63\\
\muse{} & 82 & 167 & 198 & 97.58 & 106.2\\
\meta{} & 94 & 191 & 244 & 14.28 & 16.93\\ % xia's 244 14.28 16.93
%\meta{} & 92 & 185 & 231 & 15.2 & 17.73\\ mine 
\hyb{} & 132 & 227 & 268 & 17.98 & 23.24\\
  %\pkbasic{} & 114 & 199 & 243 & 23.62 & 27.67\\
  %\pkag{} & 136 & 242 & 269 & 18.06 & 22.6\\
  \hline
  \ourtool{} & 161 & 255 & 286 & 9.48 & 14.37\\ 
  %\ourtool{}$_{PG}$  & 179 & 251 & 288 & 10.44 & 14.83\\
  %\ourtool{}$_{PGA}$ & 185 & 264 & 295 & 9.04 & 13.73\\
\hline
    \end{tabular}
    	\end{adjustbox}
     
\end{table}

\subsection{RQ1: Effectiveness of \ourtool{}} \label{sec:rq1}
To answer this RQ, we first present the overall \fl{} results of \ourtool{} and state-of-the-art \sbfl{} and \mbfl{} techniques on \olddefects{} in Table~\ref{table:cmp}.
Column ``Tech Name'' represents the corresponding techniques and the other columns present the results in terms of \topone{}, \topthree{}, \topfive{}, \mfr{} and \mar{}. From the table, we observe that \ourtool{} significantly outperforms all the existing techniques in terms of all the five metrics. 
For example, the \topone{} value of \ourtool{} is \tOneRep{}, 29 more than \hyb, 44 more than aggregation-based \sbfl{}, 67 more than \meta{}, and 79 more than \muse{}. 
In addition, \mar{} and \mfr{} values are also significantly improved (e.g., at least \mfrImrovement \ improvements for \mfr{} compared with all existing techniques), indicating a consistent improvement for all buggy elements in the ranked lists. Note that we observe that \sbfl{} outperforms state-of-the-art \mbfl{} techniques in terms of Top-ranked bugs, which is not consistent with prior \fl{} work at the method level~\cite{li2017transforming}. We find the main reason to be that the prior work did not use suspicious aggregation (proposed in parallel with the prior work) for \sbfl{}. This further demonstrates the effectiveness of suspiciousness aggregation for \sbfl. 

\begin{figure}[h] 
    \centering 
    \includegraphics[scale=0.33]{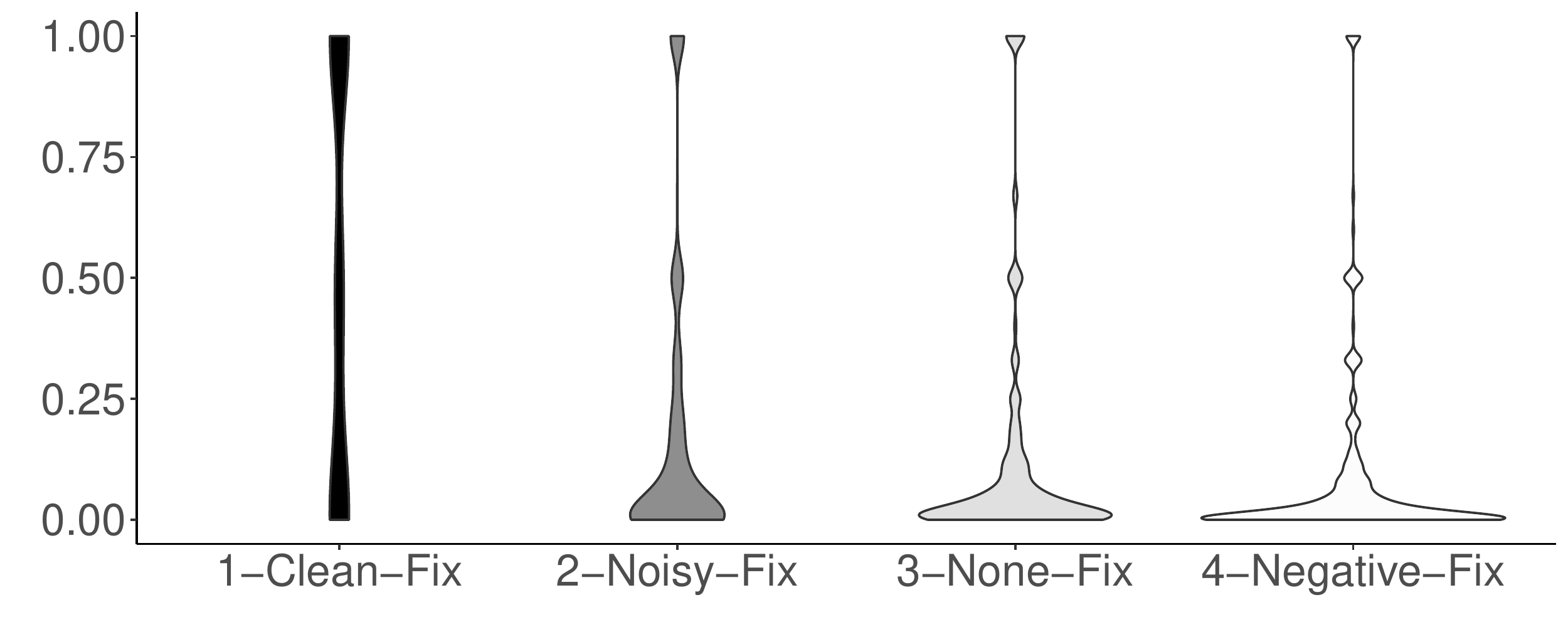}
    \caption{$\faultratio{}$ distribution for different patch groups}
    \label{fig:ratio-violin}
\end{figure}

To further investigate why the simple \ourtool{} approach works, we further analyze each of the four basic \ourtool{} patch categories in a post-hoc way. For each patch category group $\groupi$, for each bug in the benchmark, we use metric $\faultratio{}$ to represent the ratio of the number of buggy elements (i.e., methods in this work) categorized into group $\groupi$ to the number of all elements categorized into group $\groupi$. Formally, it can be presented as:
\begin{equation} \label{eq:ratio}
    \faultratio(\groupi{}) = \frac{|\{\meth{}| \group{\patch}= \groupi \wedge \patch \in \patchLoc{\meth}\} \wedge \meth{} \in \bugs{}|}
    {|\{\meth{}| \group{\patch}= \groupi \wedge \patch \in \patchLoc{\meth}\} |}
\end{equation}
where $\bugs{}$ represents a set of buggy elements. $\faultratio{}$ ranges from 0 to 1, and a higher value indicates a higher probability for a patch group to contain the actual buggy element(s). We present the distribution of the $\faultratio{}$ values on all bugs for each of the four different patch groups in the violin plot in Figure~\ref{fig:ratio-violin}, where the \emph{x} axis presents the four different groups, the \emph{y} axis presents the actual $\faultratio{}$ values, and the width of each plot shows the distribution's density. From the figure we observe that the four different groups have totally different $\faultratio{}$ distributions. E.g., group \patchoneS{} has rather even distribution, indicating that roughly half of the code elements within this group could be buggy; on the contrary, group \patchfourS{} mostly have small $\faultratio{}$ values, indicating that elements within this group are mostly not buggy. Such group analysis further confirms our hypothesis that different patch categories can be leveraged as the feedback information for powerful fault localization.
\Comment{Firstly, we find the distribution of the each two finer patch categories is similar, i.e., \smallpatchoneL{} with \smallpatchtwoL, and \smallpatchthreeL{} with \smallpatchfourL{}.
It explains why the extended four rules based on the these four finer categories show no difference in effectiveness, since the each two of them have no difference of the capability of distinguishing out the bug element.
Secondly, we observe a obvious different distribution among the patch categories defined at coarse level (i.e., in Definition~\ref{def:patch1} to Definition~\ref{def:patch4}), which also demonstrates the categorization aggregation strategy  of \ourtool{} is reasonable.   }

\Comment{The results indicate there is a really huge potential for utilizing \fin{} from program repair to improve \fl{}, since such simplistic feedback strategy can bring surprisingly large benefits. }

\Comment{To further find out, in what of kind of repair cases, \ourtool{} is effective, we categorize all the bugs into three types:
(1) \bugone{}: the bugs for which plausible patches can be generated;
(2) \bugtwo{}: the bugs for which although no plausible patch can be generated by at least one failed test can be fixed;
(3) \bugthree{}: the rest bugs for which none of failed tests can be fixed.
We also present the \fl{} results of these techniques on three bug sets in Table~\ref{table:cmp} with each row respectively named "\bugone{}, \bugtwo{}, \bugthree{}".

We have following observations on \ourtool{}.
Firstly, \ourtool{} keeps effective in all these three bug sets, which indicates the general program cases, \ourtool{} can conduct effective \fl{} with the feedback information from patch execution results.
Secondly, \ourtool{} shows a more effective performance in \bugone{} and \bugtwo{} than \bugthree, which indicates the stronger benefits from the fixing information.
In other words, in the cases where \apr{} can fix some failed test, the feedback can be more effective.}

\Comment{One potential reason might be that, \muse{} relies on the test output change (i.e., fail to pass/pass to fail) to calculate the suspiciousness value for each element, which is sufficient and precise especially in the cases where plausible patch can be generated.
That also explains why \meta{} is less effective than \muse{} in \bugone{}, since besides test output, \meta{} also takes stack-trace, exception message, and type as input, which are often too sensitive and easy to disturb the effectiveness of test output.}

\begin{tcolorbox}[boxsep=\mybox]
\textbf{Finding 1:} Simplistic feedback information from program repair can significantly boost existing \sbfl{}-based fault localization techniques, opening a new dimension for fault localization via program repair.
\end{tcolorbox}
\vspace{-3mm}

\begin{figure}[htbp]
\centering
\subfigure[Top-1]{
\includegraphics[scale=0.15]{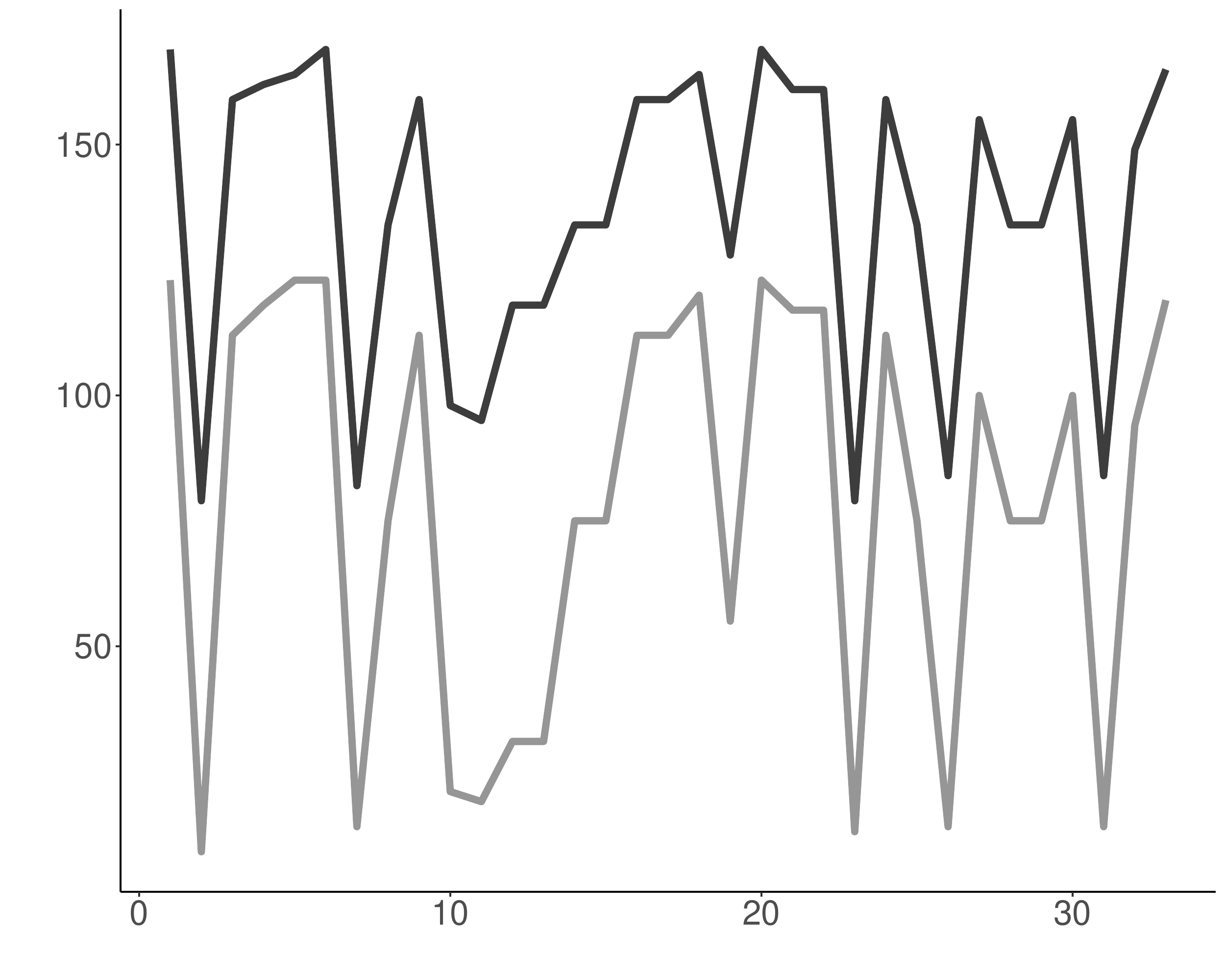}
%width=7cm, height=2.5cm
%\caption{fig1}
}
\quad
\subfigure[MAR]{
\includegraphics[scale=0.15]{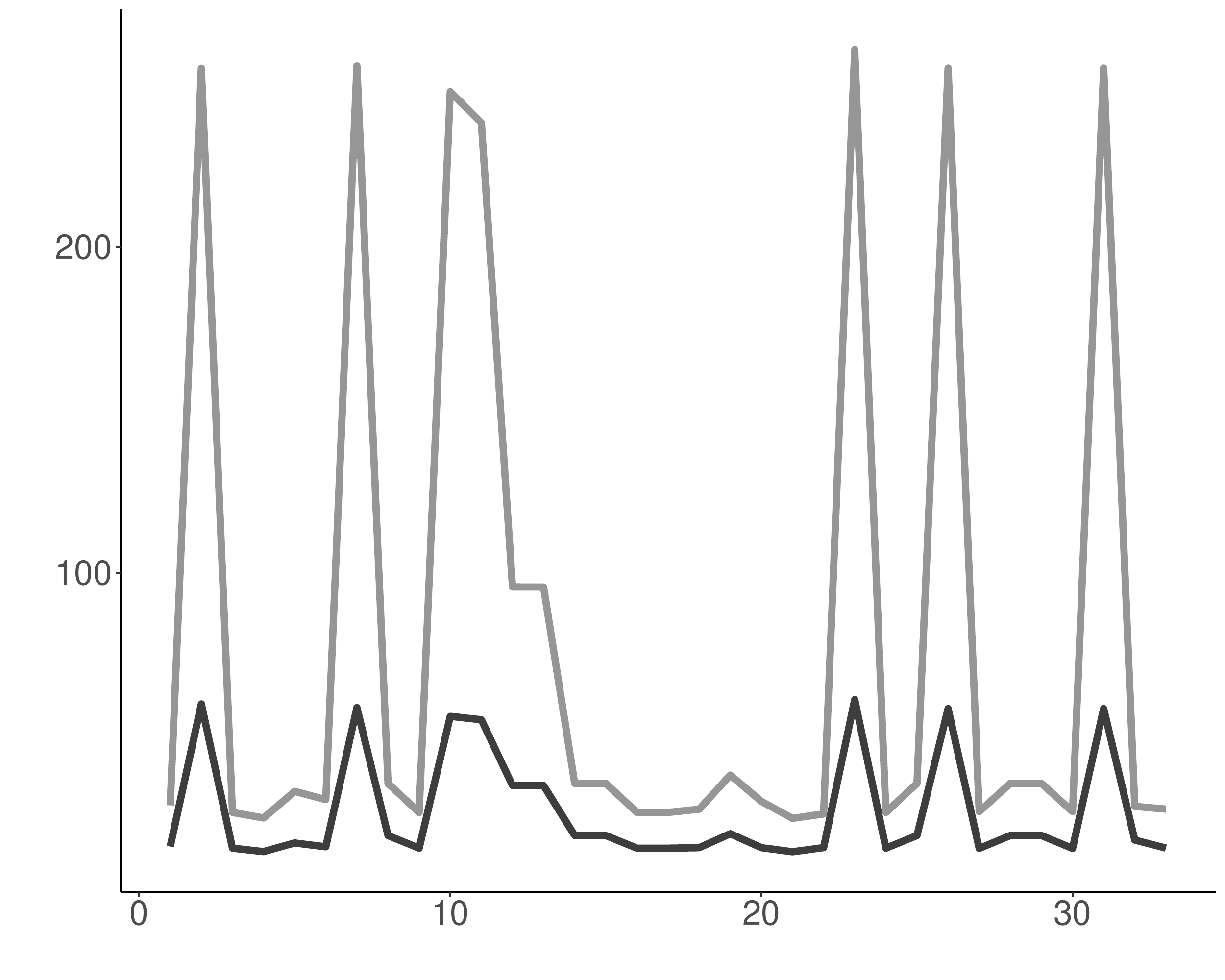}
}

\caption{Comparison of \ourtool{} and \sbfl{} over all formulae}\label{fig:spe:all}
\end{figure}

\subsection{RQ2: Different experimental configurations}\label{sec:rq2}

%%RQ2a table
\begin{table}
	\centering
	
		\caption{Impacts of finer patch categorization}\label{table:res-rules}
		\begin{adjustbox}{width=0.9\columnwidth}
	\begin{tabular}{l|l|l|l|l|l}
	\hhline
	Tech & \topone{} & \topthree{} & \topfive{} & \mfr{} & \mar{}  \\
	
	\cline{1-6}
	%%%%%%%%%%%%%%%%%
    \ourtool{} & 161 & 255 & 286 & 9.48 & 14.37 \\
    \hline
	%\cline{7-8}
\varone{} & 162 & 255 & 286 & 9.53 ($p$=0.974) & 14.41 ($p$=0.933)  \\
\vartwo{} & 161 & 252 & 283 & 9.56 ($p$=0.904) & 14.45 ($p$=0.876) \\
\varthree{} & 161 & 255 & 285 & 9.67 ($p$=0.987)& 14.62 ($p$=0.899) \\
\varfour{} & 162 & 251 & 285 & 9.55 ($p$=0.949) & 14.45 ($p$=0.967) \\
	%%%%%%%%%%%%%%%%%
	\hhline
	\end{tabular}
	
        \end{adjustbox}
\end{table}

\subsubsection{RQ2a: Impact of finer categorization}\label{sec:rq2a}
To investigate the four extended rules on the finer categorization presented in Section~\ref{sec:variant}, we implemented different \ourtool{} variants based on each rule in Table~\ref{table:variants}.
The experimental results for all the variants are shown in Table~\ref{table:res-rules}. In the table, Column ``Tech'' presents each of the compared variants and the remaining columns present the corresponding metric values computed for each variant. Note that the four variants of \ourtool{} implemented with different rules shown in Table~\ref{table:variants} are denoted as \varone{}, \vartwo{}, \varthree{} and \varfour{}, respectively.
From the table, we observe that \ourtool{} variants with  different extended rules perform similarly with the default setting in all the used metrics. To confirm our observation, we further perform the Wilcoxon signed-rank test~\cite{wiki:Wilcoxon} (at the significance level of 0.05) to compare each variant against the default setting in terms of both the first and average buggy-method ranking for each bug. The test results are presented in the parentheses in the \mfr{} and \mar{} columns, and show that there is no significant difference (p>>0.05) among the compared variants, indicating that considering the finer-grained grouping does not help much in practice.

\Comment{check whether there is significant difference among the effectiveness of these rules.
We conduct two statistic analysis for groups: (1) \apbasic{}, \varone{} and \vartwo{}, which shows the impact of finer categorization on $\patchoneS{}$; (2) \apbasic{}, \varthree{} and \varfour{}, which shows the impact of finer categorization on $\patchtwoS{}$.
Since all the data are not satisfying normal distribution,  we conduct non-parametric one-way analysis , Kruskal–Wallis~\cite{mckight2010kruskal}, for both group, and present the results in Table~\ref{table:kw}.
It shows that there is no significant difference among both groups, which indicates the finer patch categorization has no impact on the effectiveness.
In summary, we find the patch categorization and the aggregation strategy in layer-2 and layer-3 are reasonable and explainable;
for the finer patch categorization and the aggregation strategies, they have no impact on the effectiveness.}

\begin{tcolorbox}[boxsep=\mybox]
\textbf{Finding 2:} Finer-grained patch grouping has no significant impact on \ourtool{}, further demonstrating the effectiveness of the default grouping.
\end{tcolorbox}
 \vspace{-4mm}

\subsubsection{RQ2b: Impact of SBFL formulae}\label{sec:rq2b}
Our \ourtool{} approach is general and can be applied to any \sbfl{} formula, therefore, in this RQ, we further study the impact of different \sbfl{} formulae on \ourtool{} effectiveness. The experimental results are shown in Figure~\ref{fig:spe:all}. In this figure, the \emph{x} axis presents all the 34 \sbfl{} formulae considered in this work, the \emph{y} axis presents the actual metric values in terms of \topone{} and \mar{}, while the light and dark lines represent the original \sbfl{} techniques and our \ourtool{} version respectively. We can observe that, for all the studied \sbfl{} formulae, \ourtool{} can consistently improve their effectiveness. 
For example, the \topone{} improvements range from 41 (for ER1a) to 87 (for GP13), while the \mar{} improvements range from 36.54\% (for Wong) to 77.41\% (for GP02). Other metrics follow similar trend, e.g., the improvements in \mfr{} are even larger than \mar{}, ranging from 49.24\% (for SBI) to  80.47\% (for GP02). Furthermore, besides the consistent improvement, we also observe that the overall performance of \ourtool{} is quite stable for different \sbfl{} formulae. For example, the \mar{} value for \sbfl{} has huge variations when using different formulae, while \ourtool{} has stable performance regardless of the formula used, indicating that \ourtool{} can boost ineffective \sbfl{} formulae even more. 

\begin{tcolorbox}[boxsep=\mybox]
\textbf{Finding 3:} \ourtool{} can consistently improve all the \numsbfl{} studied \sbfl{} formulae, e.g., by 49.24\% to  80.47\% in MFR.
\end{tcolorbox}
 \vspace{-4mm}

%%%%%%%%%%%%%%%%%Rq2c table mutant%%%%%%%%%%%%%%%%%%%%
\begin{table}
	\centering
	\caption{Impacts of using mutation or repair information}\label{table:mutant}
		\begin{adjustbox}{width=0.85\columnwidth}
	\begin{tabular}{l|l|l|l|l|l}

	\hhline
	Tech Name & \topone{} & \topthree{} & \topfive{} & \mfr{} & \mar{} \\ \hhline
		
	%%%%%%%%%%%%%%%%%
%	\muse{}$_{PIT}$ & 86 & 152 & 183 & 46.89 & 52.1 \\ mine
    \muse{}$_{PIT}$ & 82 & 167 & 198 & 97.58 & 106.2 \\ %xia
	\muse{}$_{PraPR}$ & 95 & 172 & 207 & 38.79 & 43.1\\ \hline
	%\muse{} & 82 & 167 & 198 & 97.58 & 106.2\\

   % \meta{}$_{PIT}$ & 92 & 185 & 231 & 15.2 & 17.73 \\ % mine
    \meta{}$_{PIT}$ & 94 & 191 & 244 & 14.28 & 16.93\\ %xia
    \meta{}$_{PraPR}$ & 77 & 170 & 211 & 21.42 & 22.94\\ \hline
        
    \hyb{}$_{PIT}$ & 132 & 227 & 268 & 17.98 & 23.24\\
    \hyb{}$_{PraPR}$ & 130 & 228 & 267 & 18.03 & 23.28\\ \hline

    \ourtool{}$_{PIT}$ & \tOneMut{} & 238 & 266 & 15.24 & 20.33\\
    \ourtool{}$_{PraPR}$ & \tOneRep{} & 255 & 286 & 9.48 & 14.37\\

	%%%%%%%%%%%%%%%%%
		
	\hhline
	\end{tabular}
	\end{adjustbox}

\end{table}

\subsubsection{RQ2c: Impact of feedback source}\label{sec:rq2c}
Since \ourtool{} is general and can even take traditional mutation testing information as feedback source, we implement a new \ourtool{} variant that directly take mutation information (computed by PIT) as feedback. To distinguish the two \ourtool{} variants, we denote the new variant as \ourtool$_{PIT}$ and the default one as \ourtool$_{\prapr}$. Meanwhile, all the existing \mbfl{} techniques can also take the \apr{} results from \prapr{} as input (\prapr{} can be treated as an augmented mutation testing tool with more and advanced mutators), thus we also implemented such variants for traditional \mbfl{} for fair comparison, e.g., the original \muse{} is denoted as \muse{}$_{PIT}$ while the new \muse{} variant is denoted as \muse$_{\prapr}$. Table~\ref{table:mutant} presents the experimental results for both \ourtool{} and prior mutation-based techniques using different information sources. We have the following observations:

First, \ourtool{} is still the most effective technique compared with other techniques even with the feedback information from mutation testing. For example, \ourtool{} with mutation information localizes 141 bugs within \topone{}, while the most effective existing technique (no matter using mutation or repair information) only localizes 132 bugs within \topone{}.
This observation implies that the \ourtool{} approach of using feedback information (from program-variant execution) to refine \sbfl{} ranking is general in design, and is not coupled tightly with specific source(s) of feedback.

Second, \ourtool{} performs worse when feedback source changes from program repair to mutation testing.
For example, the \topone{} decreases from 161 to 141. The reason is that patches within groups \patchoneS{}/\patchtwoS{} can help promote the ranking of buggy methods. However, mutation testing cannot create many such patches. For example, we find that the number of bugs with \patchoneS{}/\patchtwoS{} patches increase by 40.0\% when changing from mutation testing to \apr{}. This further indicates that \emph{\apr{} is more suitable than mutation testing for fault localization since it aims to pass more tests while mutation testing was originally proposed to fail more tests.}

Third, for the two existing \mbfl{} techniques, \muse{} performs better in program repair compared to mutation testing while \meta{} is the opposite. We find the reason to be that \muse{} simply counts the number of tests changed from passed to failed and vice versa, while \meta{} leverages the detailed test failure messages to determine mutant impacts. In this way, \apr{} techniques that make more failed tests pass can clearly enhance the results of \muse, but do not have clear benefits for \meta{}.

\Comment{The reason might be that, \muse{} is also more effective when there is more information on the cases failed test can fixed,  while \meta{} takes too much information as input and is not sensitive to the change in test output.
}
\begin{tcolorbox}[boxsep=\mybox]
\textbf{Finding 4:} \ourtool{} still performs well even with the mutation feedback information, but has effectiveness decrements compared to using program repair, indicating the superiority of program repair over mutation testing for fault localization.
\end{tcolorbox}
 \vspace{-4mm}

%%%%%%%%%%%Table for rq2d partial
\begin{table}
	\centering
	\caption{Impacts of using partial matrices}\label{table:part}
	\begin{adjustbox}{width=1\columnwidth}
	\begin{tabular}{l|l|l|l|l|l|l}

\hhline
$\pMatrixpart{}$  & Tech Name & \topone{} & \topthree{} & \topfive{} & \mfr{} & \mar{} \\ \hline
		
%%%%%%%%%%%%%%%%%
\multirow{3}{0.3in}{\orderone{}}
 & \muse{}$_{\prapr}$ & 92 & 148 & 172 & 118.56 & 125.11\\
 & \meta{}$_{\prapr}$ & 64 & 128 & 167 & 113.9 & 126.79\\
 & \ourtool{} & 165 & 254 & 287 & 15.46 & 20.96\\
\hline
\multirow{3}{0.3in}{\ordertwo{}}
 & \muse{}$_{\prapr}$ & 87 & 130 & 152 & 191.71 & 206.0\\
 & \meta{}$_{\prapr}$ & 32 & 73 & 94 & 163.29 & 170.61\\
 & \ourtool{} & 169 & 252 & 285 & 9.17 & 15.15\\
\hline
\multirow{3}{0.3in}{\orderthree{}}
 & \muse{}$_{\prapr}$ & 89 & 128 & 144 & 169.33 & 174.09\\
 & \meta{}$_{\prapr}$ & 63 & 127 & 159 & 187.19 & 195.07\\
 %& \hyb{} & 131 & 228 & 267 & 18.55 & 24.09\\
 & \ourtool{} & 158 & 244 & 278 & 19.07 & 25.25\\

	\hhline
	\end{tabular}
        \end{adjustbox}
\end{table}

\subsubsection{RQ2d: Impact of partial execution matrix}\label{sec:rq2d}
So far, we have studied \ourtool{} using full patch execution matrices. However, in practical program repair, a patch will not be executed against the remaining tests as soon as some test falsifies it for the sake of efficiency. Therefore, we further study new \ourtool{} variants with only partial patch execution matrices. 
The experimental results for three variants of \ourtool{} using different partial matrices are shown in Table~\ref{table:part}. From the table, we have the following observations:

First, surprisingly, \ourtool{} with different partial matrices still perform similarly with our default \ourtool{} using full matrices, while the traditional \mbfl{} techniques perform significantly worse using partial matrices. We think the reason to be that existing \mbfl{} techniques utilize the detailed number of impacted tests for fault localization and may be too sensitive when switching to partial matrices.
\Comment{
The table shows that, with partial matrix information, \ourtool{} is still the most effective technique compared to the other  at all metrics, such as \topn{}, \mar{} and \mfr{}, which indicates for all the bug elements, \ourtool{} is still effective to localize them. }
Second, \ourtool{} shows consistent effectiveness with partial matrices obtained from different test execution orderings, e.g., even the worst ordering still produces 158 \topone{} bugs. One potential reason that \orderthree{} performs the worst is that if there is any passed tests changed into failing, the original failed tests will no longer be executed, missing the potential opportunities to have \patchoneS{}/\patchtwoS{} patches that can greatly boost fault localization. Luckily, in practice, repair tools always execute the failed tests first (i.e., \ordertwo{}), further demonstrating that \ourtool{} is practical. 
\Comment{We find that \orderone{} which is the default order appears stable in \topn{} but become less effective in \mar{} and \mfr{} compared to full matrix; \ordertwo{} which executes the failed tests first even become little better; while \orderthree{} which execute the pass test first become little worse.
The potential reason might be that, the failed test execution is more essential to \ourtool{}, while executing passing tests first would lose such information. 
\ordertwo{} is exactly the common practice in program repair scenario, which shows \ourtool{} is useful in practice.}

We next present the cost reduction benefits that partial execution matrices can bring to speed up the \ourtool{} fault localization process. The experimental results for the HEAD version (i.e., the latest and usually the
largest version) of each studied subject are shown in Table~\ref{table:time}. In the table, Column ``\fulltime{}'' presents the time for executing all tests on each candidate patch, Column ``\parttime{}'' presents the time for terminating test execution on a patch as soon as the patch gets falsified (following the default test execution order of \prapr{}, i.e., executing originally failed tests first then passed tests), Columns ``Reduced Time" and ``Reduced Ratio'' show the reduced time and the reduction ratio from \fulltime{} to \parttime{}. We use 4 threads for executing both \prapr{} variants.
From the table, we can observe that partial execution matrix collection can overall achieve \exeOverheadGain{} reduction compared to full matrix collection. Furthermore, using partial execution matrices, even the largest Closure subject only needs less than 2 hours, indicating that \ourtool{} can be scalable to real-world systems (since we have shown that \ourtool{} does not have effectiveness drop when using only partial matrices).
 \vspace{-2mm}
\begin{tcolorbox}[boxsep=\mybox]
\textbf{Finding 5:} \ourtool{} keeps its high effectiveness even on partial patch execution matrices, especially with test execution ordering following the program repair practice, demonstrating that its overhead can be reduced by \exeOverheadGain{} without clear effectiveness drop.
\end{tcolorbox}
 \vspace{-4mm}

\begin{table}
	\centering

		\caption{Collection time for full and partial matrices}\label{table:time}
		\begin{adjustbox}{width=0.9\columnwidth}
	\begin{tabular}{l|l|l|l|l}
		\hhline

		\textbf{Subject} & \fulltime{} & \parttime{} & Reduced Time & Reduced Ratio \\ 	\hhline
		
Lang-1 & 0m38s & 0m31s & 0m7s & 18.4\% \\ 
%\hline
Closure-1 & 2568m26s & 110m33s & 2457m53s & 95.7\% \\ 
%\hline
Mockito-1 & 452m33s & 2m43s & 449m50s & 99.4\% \\ 
%\hline
Chart-1 & 32m27s & 2m41s & 29m46s & 91.7\% \\ 
%\hline
Time-1 & 149m14s & 0m41s & 148m33s & 99.5\% \\ 
%\hline
Math-1 & 68m24s & 7m53s & 60m31s &88.5\% \\ \hline
\textbf{Total} & 3271m42s & 125m2s & 3146m40 &96.2\% \\ \hline
		
\hline
\end{tabular}
\end{adjustbox}

\end{table}

\begin{table}
	
	\caption{Results on \newdefects{}}\label{table:newbench}

	\centering
	\begin{adjustbox}{width=0.7\columnwidth}
	\begin{tabular}{l|l|l|l|l|l}

	\hhline
	 Tech Name & \topone{} & \topthree{} & \topfive{} & \mfr{} & \mar{} \\ \hline
		
	%%%%%%%%%%%%%%%%%
	\sbfl{} & 59 & 102 & 124 & 13.81 & 20.44\\
    \muse{} & 42 & 75 & 82 & 53.97 & 60.17\\
    \meta{} & 43 & 80 & 102 & 19.05 & 24.9\\
    \hyb{} & 65 & 110 & 130 & 13.28 & 19.88\\
    \hline
    \ourtool{} & 78 & 117 & 131 & 12.01 & 17.96\\

	%%%%%%%%%%%%%%%%%
	\hhline
	\end{tabular}
	\end{adjustbox}
\end{table}

\subsubsection{RQ2e: Impact of used benchmarks}\label{sec:rq2e}

In this RQ, we further compare \ourtool{} and state-of-the-art \sbfl{}/\mbfl{} techniques on additional bugs from \newdefects{}, to reduce the threats to external validity. The experimental results are shown in Table~\ref{table:newbench}. From the table, we observe that \ourtool{} still significantly outperforms all other compared techniques. 
E.g., \topone{} is improved from 59 to 78 compared to the original state-of-the-art \sbfl. Such a consistent finding on additional bugs further confirms our findings in RQ1.
\Comment{We further present the \topone{} results of each technique on each subject, to make sure the overall improvement is not dominated by certain subject in  Figure~\ref{fig:radar}.
The Radar plot shows that almost on each subject, \ourtool{} is most effective, which sufficiently shows the generality of the effectiveness of \ourtool{}.
}

\begin{tcolorbox}[boxsep=\mybox]
\textbf{Finding 6:} \ourtool{} still significantly outperforms state-of-the-art \sbfl{} and \mbfl{} on additional bugs.
\end{tcolorbox}
 \vspace{-4mm}

\subsection{RQ3: Boosting learning-based localization}

We further apply the basic \ourtool{} to boost state-of-the-art unsupervised-learning-based (i.e., \pkbasic{} and \pkag{}~\cite{zhang2019empirical}) and supervised-learning-based (i.e., \dpfl{}~\cite{li2019deepfl}) fault localization. For unsupervised-learning-based techniques, \ourtool{} is generic and can use any existing fault localization techniques to compute initial suspiciousness (Section~\ref{sec:basic}); therefore, we directly apply \ourtool{} on the initial suspiciousness computed by \pkbasic{} and \pkag{}, denoted as \ourtool{}$_{\pkbasic{}}$ and  \ourtool{}$_{\pkbasic{}MA}$, respectively. For supervised-learning-based techniques, \ourtool{} with all the 34 used \sbfl{} formulae can serve as an additional feature dimension; therefore, we augment \dpfl{} by injecting \ourtool{} features between the original mutation and spectrum feature dimensions (since they are all dynamic features), and denote that as \ourtool{}$_{\dpfl{}}$. The experimental results are shown in Table~\ref{table:pgk}. Note that \dpfl{} results are averaged over 10 runs due to the DNN randomness~\cite{li2019deepfl}. First, even the basic \ourtool{} significantly outperforms state-of-the-art unsupervised-learning-based fault localization. E.g., \ourtool{} localizes 161 bugs within \topone{}, while the most effective unsupervised \pkag{} only localizes 136 bugs within \topone{}. Second, \ourtool{} can significantly boost unsupervised-learning-based fault localization. E.g., \ourtool{}$_{\pkbasic{}MA}$ localizes 185 bugs within \topone{}, \emph{the best fault localization results on \defects{} without supervised learning to our knowledge}. Actually, such unsupervised-learning-based fault localization results even significantly outperform many state-of-the-art supervised-learning-based techniques, e.g., TraPT~\cite{li2017transforming}, FLUCCS~\cite{sohn2017fluccs}, and CombineFL~\cite{zou2019empirical} only localize 156, 160, and 168 bugs from the same dataset within \topone{}, respectively~\cite{zou2019empirical, li2019deepfl}. Lastly, we can observe that \ourtool{} even boosts state-of-the-art supervised-learning-based technique. E.g., it boosts \dpfl{} to localize 216.8 bugs within \topone{}, \emph{the best fault localization results on \defects{} with supervised learning to our knowledge}.

\Comment{the advanced unsupervised-learning-based \fl{} technique to investigate whether \ourtool{} can further boost the effectiveness of  unsupervised-learning \fl{} in Table~\ref{table:pgk}.
We observe a huge improvement for \ourtool{} on the unsupervised-learning-based techniques.
For \pkbasic{}, \ourtool{} improves \topone{} from 114 to 179 with 57\% improvement and for \pkag{}, the \topone{} is 185, which indicates the effectiveness of \ourtool{} over unsupervised-learning techniques.
}
 \vspace{-0.8mm}
\begin{tcolorbox}[boxsep=\mybox]
\textbf{Finding 7:} \ourtool{} significantly outperforms state-of-the-art unsupervised-learning-based fault localization, and can further boost unsupervised and supervised learning based fault localization, further demonstrating the effectiveness and general applicability of \ourtool{}.
\end{tcolorbox}
 \vspace{-1mm}

%%%%%%%%%%%%%%%%%Rq2c table mutant%%%%%%%%%%%%%%%%%%%%
\begin{table}[h!]
	\centering
	\caption{Boosting state-of-the-art \pkbasic{}}\label{table:pgk}
	\begin{adjustbox}{width=0.8\columnwidth}
	\begin{tabular}{l|l|l|l|l|l}

	\hhline
	Tech Name & \topone{} & \topthree{} & \topfive{} & \mfr{} & \mar{} \\ \hhline
		
	%%%%%%%%%%%%%%%%%
  \pkbasic{} & 114 & 199 & 243 & 23.62 & 27.67\\
    \ourtool{}$_{\pkbasic{}}$  & 179 & 251 & 288 & 10.44 & 14.83\\ \hline
  \pkag{} & 136 & 242 & 269 & 18.06 & 22.6\\
    \ourtool{}$_{PRFL_{MA}}$ & 185 & 264 & 295 & 9.04 & 13.73\\ \hline 
     \dpfl{} & 211.0&	284.5 &	310.5 &	4.97&	6.27 \\
     \ourtool{}$_{\dpfl{}}$ & 216.8 & 293.6 & 318.0 & 4.53 &5.88 \\ 
    
	%%%%%%%%%%%%%%%%%
	\hhline
	\end{tabular}
	\end{adjustbox}

\end{table}

\section{Conclusion}
\label{sec:conclude}
We have investigated a simple question: \emph{can automated program repair help with fault localization?} To this end, we have designed, \ourtool{}, the first approach that leverages program repair information as the feedback for powerful fault localization. The experimental results on the widely used \defects{} benchmarks demonstrate that \ourtool{} can significantly outperform state-of-the-art spectrum and mutation based fault localization. Furthermore, we have demonstrated \ourtool{}'s effectiveness under various settings. Lastly, \ourtool{} even boosts state-of-the-art fault localization via both unsupervised and supervised learning. In the near future, we will work on \emph{tentative program repair}, a new direction enabled by this research to allow fault localization and program repair to boost each other for more powerful debugging, e.g., patch execution results from an initial program-fixing template set can enable precise fault localization for applying later more advanced program-fixing template set(s) later for cost-effective debugging.

\balance
\bibliographystyle{ACM-Reference-Format}
\bibliography{ref,sigproc,simpr}

\end{document}